\documentclass[a4paper,11pt]{article}
\pdfoutput=1 
\usepackage{amssymb}
\usepackage{amsmath}
\usepackage{graphicx}
\usepackage{caption}

\usepackage{amsthm}
\newtheorem{prop}{Proposition}

\begin{document}
\thispagestyle{empty}

\begin{flushright}
ITFA-2008-09\\
DIAS-STP-08-04\\
EMPG-08-02\\
\end{flushright}

\begin{center}
\baselineskip 24 pt {\Large \bf Magnetic Charge Lattices, 
Moduli Spaces and Fusion Rules}

\vspace{0.2cm}

\baselineskip 16 pt

\vspace{.7cm} 
{\large L.~Kampmeijer}\footnote{leo.kampmeijer@uva.nl}$^{,a}$,
{\large J.~K.~Slingerland}\footnote{slingerland@stp.dias.ie}$^{,b}$,
{\large  B.~J.~Schroers}\footnote{bernd@ma.hw.ac.uk}$^{,c}$,
{\large F.~A.~Bais}\footnote{bais@science.uva.nl  }$^{,a}$\\

\vspace{0.5cm}
\emph{
{}$^a$ Institute for Theoretical Physics, University of Amsterdam,
 \\  Valckenierstraat 65, 1018 XE Amsterdam, The Netherlands\\
\vspace{.2cm}
$^b${ Dublin Institute for Advanced Studies, School for Theoretical Physics, }
   \\ { 10 Burlington Rd, Dublin, Ireland}\\
\vspace{.2cm}
$^c${ Department of Mathematics and Maxwell Institute for Mathematical Sciences} 
\\
{  Heriot-Watt University,
Edinburgh EH14 4AS, United Kingdom}  \\
}
\vspace{.7cm}
{March   2008}

\end{center}

\vspace{.2cm}


\begin{abstract}
We analyze the set of magnetic charges carried by smooth BPS monopoles in Yang-Mills-Higgs theory with arbitrary gauge group $G$ spontaneously broken to a subgroup $H$. The charges are restricted by a generalized Dirac quantization condition and by an inequality due to Murray. Geometrically, the set of allowed charges is a solid cone in the coroot lattice of $G$, which we call the Murray cone. We argue that magnetic charge sectors correspond to points in the Murray cone divided by the Weyl group of $H$; hence magnetic charge sectors are labelled by dominant integral weights of the dual group $H^*$. We define generators of the Murray cone modulo Weyl group, and interpret the monopoles in the associated magnetic charge sectors as basic; monopoles in sectors with decomposable charges are interpreted as composite  configurations. This interpretation is supported by the dimensionality of the moduli spaces associated to the magnetic charges and by classical fusion properties for smooth monopoles in particular cases. Throughout the paper we compare our findings with corresponding results for singular monopoles recently obtained by Kapustin and Witten.
\end{abstract}

 \vspace{.7cm}

\centerline{PACS numbers: 11.15.Ex, 11.15.Kc, 14.80.Hv}

\newpage
\tableofcontents
\section{Introduction}  
In the 1970s  Goddard, Nuyts and Olive were the first to write down a rough version of what has become one of the most celebrated dualities in high energy physics \cite{Goddard:1976qe}.  By generalizing the Dirac quantization condition they showed that the charges of monopoles 
take values in the weight lattice of the dual gauge group, now known as the GNO or Langlands dual group. Based on this fact they came up with a bold yet attractive conjecture: monopoles transform as representations of the dual group.\\
Within a year Montonen and Olive observed that the Bogomolny Prasad Sommerfield (BPS) mass formula for dyons \cite{Bogomolny:1975de,Prasad:1975kr} is invariant under the interchange of electric and magnetic quantum numbers if the coupling constant is inverted as well \cite{Montonen:1977sn}.  This led to the dramatic conjecture that the strong coupling regime of some suitable quantum field theory is described by a weakly coupled theory with a similar Lagrangian but with the gauge group replaced by the GNO dual group and the coupling constant inverted. Moreover they proposed that in the BPS limit of a gauge theory where the gauge group is spontaneously broken to $U(1)$ the 't Hooft-Polyakov solutions \cite{'tHooft:1974qc,Polyakov:1974ek} in the original theory correspond to the heavy gauge bosons of the dual theory. Supporting evidence for the idea of viewing  the 't Hooft-Polyakov monopoles as fundamental particles came from Erick Weinberg's zero-mode analysis in \cite{Weinberg:1979ma}.\\ 
Soon after Montonen and Olive proposed their duality, Osborn noted that $\mathcal N\!=\!4$ Super Yang-Mills theory (SYM) would be a good candidate to possess the duality since BPS monopoles fall into the same BPS supermultiplets as the elementary particles of the theory \cite{Osborn:1979tq}. $\mathcal N\!=\! 2$ SYM on the other hand has always been considered an unlikely candidate because the BPS monopoles fall into BPS multiplets that do not correspond to the elementary fields of the $\mathcal N\!=\! 2$ Lagrangian. In particular there are no semi-classical monopole states with spin equal to 1 so that the monopoles cannot be identified with heavy gauge bosons.
Most surprisingly however the Montonen-Olive conjecture has never been proven for $\mathcal N=4$ SYM whereas a different version of the duality has explicitly been shown to occur for the $\mathcal N\!=\!2$ theory in 1994 by Seiberg and Witten. They started out from $\mathcal N\!=\! 2$ SYM with the $SU(2)$ gauge group broken down to $U(1)$ \cite{Seiberg:1994rs} and computed the exact effective Lagrangian of the theory to find a strong coupling phase described by SQED except that the electrons are actually magnetic monopoles. Similar results hold for  higher rank gauge groups broken down to their maximal abelian subgroups \cite{Klemm:1994qs,Argyres:1994xh}. In these cases we indeed have an explicit realization of a magnetic abelian gauge group at strong coupling.\\
\\
One might wonder whether these theories could also have non-abelian phases at strong coupling, that is a phase where the gauge group is broken down to a non-abelian subgroup. Both the classical $\mathcal N \!=\! 4$ and $\mathcal N\!=\!2$ pure SYM theories have a continuous space of ground states corresponding to the vacuum expectation value of the adjoint Higgs field. A non-abelian phase corresponds to the Higgs VEV having degenerate eigenvalues. In the $\mathcal N=4$ theory the supersymmetry is sufficient to protect the classical vacuum structure even non-perturbatively \cite{Seiberg:1988ur}. 
So the non-abelian phases manifestly realized in the classical regime must survive at strong coupling as well. 
In $\mathcal N\!=\!2$ theory the vacuum structure is changed in quite a subtle way by non-perturbative effects. In those subspaces of the quantum moduli space where a non-abelian phase might be expected there are no massless W-bosons. Instead the perturbative degrees of freedom correspond to photons and massless 
monopoles carrying abelian charges. In the best case there are some indications that a non-abelian phase may exist at strong coupling in certain $\mathcal N=2$ theories with a sufficient number of hyper multiplets \cite{Bolognesi:2002iy,Auzzi:2002um}. 
\\
Unfortunately and despite the importance of its results, Seiberg-Witten theory seems to exclude any manifest non-abelian phase which makes it impossible to study the original GNO-conjecture on the transformation properties of non-abelian monopoles. Quite recently however Witten and Kapustin have found extraordinary new evidence to support the non-abelian Montonen-Olive conjecture. This evidence was constructed in an effort to show that the mathematical concept of  the geometric Langlands correspondence arises naturally from electric-magnetic duality in physics \cite{Kapustin:2006pk}.\\
\\
The starting point for Kapustin and Witten is a twisted version of $\mathcal N\!=\!4$ gauge theory.  They identify 't Hooft operators, which create the flux of Dirac monopoles,  with Hecke operators. The labels of these operators are given by the generalized Dirac quantization rule and can up to a Weyl transformation be identified with dominant integral weights of the dual gauge group. Note that a dominant integral weight is the highest weight of a unique irreducible representation. Magnetic charges thus correspond to irreducible representations of the dual gauge group.  The moduli spaces of the singular BPS monopoles are identified with the spaces of Hecke modifications. The operation of bringing two separated  monopoles together defines a non-trivial product of the corresponding moduli spaces. The resulting space can be stratified according to its singularities. Each singular subspace is again the compactified  moduli space of a monopole related to an irreducible representation in the tensor product. The multiplicity of the BPS saturated states for each magnetic weight is found by analyzing the ground states of the quantum mechanics on the moduli space. The number of ground states given by the De Rham cohomology of the moduli space agrees with the dimension of the irreducible representation labelled by the magnetic weight. Moreover Kapustin and Witten exploited existing mathematical results on the singular cohomology of the moduli spaces to show that the products of 't Hooft operators mimic the fusion rules of the dual group. The operator product expansion (OPE) algebra of the 't Hooft operators thereby reveals the dual representations in which the monopoles transform.\\ 
\\
There is an enormous amount of evidence to support the Montonen-Olive conjecture  for the ordinary  $\mathcal N\!= \! 4$ SYM theory, see for example \cite{Sen:1994yi,Vafa:1994tf,Harvey:1995tg}.  These results which mainly concern the invariance of the spectrum do not leave much room to doubt that the strongly coupled theory can be described in terms of monopoles. However, they do not say much about the fusion rules of these monopoles. If the original GNO conjecture does indeed apply for $\mathcal N\!= \! 4$ SYM theory with residual non-abelian gauge symmetry, smooth monopoles should have properties similar to those of the singular BPS monopoles in the Kapustin-Witten setting. By the same token we claim that one can exploit these properties to find new evidence for the GNO duality in spontaneously broken theories. This paper aims to set a first step in this direction by generalizing the classical fusion rules found by Erick Weinberg for abelian BPS monopoles \cite{Weinberg:1979zt} to the non-abelian case. Our results indicate that smooth BPS monopoles are naturally labelled by integral dominant weights of the residual dual gauge group.\\
\\
The outline of this paper is as follows. In section \ref{sect:chargelat} we recapitulate the generalized Dirac quantization condition and  describe the resulting  magnetic charge lattices for both singular and smooth monopoles and their relation with the weight lattice of the dual group. In addition we review the Murray condition which restricts the allowed charges for smooth BPS monopoles to a cone in the magnetic charge lattice. Finally we introduce the fundamental Murray cone which arises by modding out the residual Weyl group. In section \ref{sect:gencharges} we determine the additive structure of the Murray cone and the fundamental Murray cone. In both cases this results in a unique set of indecomposable charges which generate the cone. For Dirac monopoles similar sets of generating charges are introduced. We show that the generators of the fundamental Murray cone generate a subring in the representation ring of the residual gauge group. In the appendix we construct an algebraic object whose representation ring is identical to to this special subring.\\
We claim that the decomposable charges for smooth BPS monopoles correspond to multi-monopole configurations built up from  basic monopoles associated to the generating charges. To support this claim we study the relevant moduli spaces in section \ref{sect:modspace}. By analyzing the dimensions of these spaces it is shown that this multi-monopole picture only holds within the fundamental Murray cone. Further evidence for these classical fusion rules is found in section \ref{sect:fusionprop} where we review to what extent classical monopole solutions can be patched together. We briefly discuss similar results for singular BPS monopoles and speculate on the implications for the semi-classical fusion rules.
\section{Magnetic charge lattices}
\label{sect:chargelat}
In this section we describe and identify the magnetic charges for several classes of monopoles. We shall start with a review for Dirac monopoles, then continue with smooth monopoles in spontaneously broken theories. Specifically for adjoint symmetry breaking we shall explain how the magnetic charge lattice can be understood in terms of the Langlands or GNO dual group of either the full gauge group or the residual gauge group. This will finally culminate in a thorough description of the set of magnetic charges for smooth BPS monopoles.\\
\\
Dirac monopoles can be described as solutions of the Yang-Mills equations with the property that they are time independent  and rotationally invariant. More importantly they are singular at a point. As a direct generalization of the Wu-Yang description of $U(1)$ monopoles \cite{Wu:1975es}, singular monopoles in Yang-Mills theory with gauge group $H$ correspond to a connection on an
$H$-bundle on a sphere surrounding the singularity.  The $H$-bundle may be topologically non-trivial, but in addition the
monopole connection equips the bundle  with a holomorphic structure.  The classification
of monopoles in terms of their magnetic charge 
then becomes equivalent to Grothendieck's classification of $H$-bundles on $\mathbb{CP}^1$.
As a result, the magnetic charge has topological and holomorphic components, both of which play an important role in 
this paper.\\
A different class of monopoles is found from smooth static solutions of a Yang-Mills-Higgs theory on $\mathbb R^3$ where the gauge group $G$ is broken to a subgroup $H$. Since $\mathbb R^3$ is contractible   the $G$-bundle is necessarily trivial. Choosing the boundary conditions so that the total energy is finite while the total magnetic charge is nonzero one finds that smooth monopoles behave asymptotically as Dirac monopoles. Since the long range gauge fields correspond to the residual gauge group this gives a non-trivial $H$-bundle at spatial infinity. The charges of smooth monopoles in a theory with $G$ spontaneously broken to $H$  are thus a subset in the magnetic charge lattice of singular monopoles in a theory with gauge group $H$ .\\
Finally one can restrict solutions to the BPS sector where the energy is minimal. This limitation is natural in supersymmetric Yang-Mills theories with a broken gauge group but with unbroken supersymmetry such that the potential vanishes identically. Smooth BPS monopoles are solutions of the BPS equations and therefore automatically solutions of the full equations of motion of the Yang-Mills-Higgs theory. Thus the charges of BPS monopoles are in principle a subset of the charges of smooth monopoles. This subset is determined by the so-called Murray condition which we shall introduce below.
\subsection{Quantization condition for singular monopoles}
\label{sect:singcondition}
The magnetic charge of a singular monopole is restricted by the generalized Dirac quantization condition \cite{Englert:1976ng,Goddard:1976qe}. This consistency condition can be derived from the bundle description \cite{Wu:1975es}. One can work in a gauge where 
 the magnetic field has the  form
\begin{equation}
B = \frac{G_0}{4\pi r^2} \hat r,
\end{equation}
with  $G_0$  an element in the Lie algebra of the gauge group $H$. This magnetic field corresponds to a gauge potential given by:
\begin{equation}
A_\pm = \pm\frac{G_0}{4\pi}\left(1\mp \cos \theta\right)d\varphi.
\end{equation}
The indices of the gauge potential refer to the two hemispheres. 
On the equator where the two patches overlap the gauge potentials are related by a gauge transformation: 
\begin{equation}
A_- = \mathcal{G}^{-1}(\varphi)\left( A_+ + \frac{i}{e}d\right) \mathcal{G}(\varphi).
\end{equation}
One can check
\begin{equation}
\label{eqn:equator}
\mathcal{G}(\varphi)= \exp\left(\frac{ie}{2\pi}G_0\varphi\right).
\end{equation} 
One obtains similar transition functions for  associated vector bundles by substituting  appropriate matrices representing $G_0$.  All such transition functions must be single valued. In the Dirac picture this means that under parallel transport around the equator  electrically charged fields should not detect the Dirac string. Consequently we find for each representation the condition:
\begin{equation}
\mathcal G(2\pi)= \exp\left(ieG_0\right)
= \mathbb I,
\end{equation} 
where $\mathbb I$ is the unit matrix. 
To cast this condition in slightly more familiar form we note that there is a gauge transformation that maps the magnetic field and hence also $G_0$ to a Cartan subalgebra (CSA) of $H$. Thus without loss of generality we can take $G_0$ to be a linear combination of the generators $\left( H_a \right)$ of the CSA in the Cartan-Weyl basis:
\begin{equation}
G_0=  \frac{4\pi}{e} \sum_a g_a\cdot H_a \equiv \frac{4\pi}{e} g\cdot H.
\end{equation} 
The generalized Dirac quantization condition can now be formulated as follows:
\begin{equation}
\label{diraccond}
2\lambda \cdot g \in \mathbb Z,  
\end{equation}
for all charges $\lambda$ in the weight lattice $\Lambda(H)$ of $H$.\\
\\
We thus see that the magnetic weight lattice $\Lambda^*(H)$ defined by the Dirac quantization condition is dual to the electric weight lattice $\Lambda(H)$. Consider for example the case where $H$ is semi-simple as well as simply connected so that the weight lattice  $\Lambda(H)$ is  generated by the fundamental weights $\{\lambda_i\}$.  Then $\Lambda^*(H)$ is generated by the simple coroots $\{\alpha_i^*=\alpha_i/\alpha_i^2\}$ which satisfy:
\begin{equation}
2\alpha_i^*\cdot\lambda_j =\frac{2\alpha_i\cdot \lambda_j}{\alpha^2_i} = \delta_{ij}.
\end{equation}
As originally observed by Goddard, Nuyts and Olive the magnetic weight lattice can be identified with the weight lattice of the GNO dual group $H^*$. For example if we take $H=SU(n)$ and define the roots of $SU(n)$ such that $\alpha^2=1$, we see that $\Lambda^*(SU(n))$ corresponds to the root lattice of $SU(n)$. The root lattice of $SU(n)$ on the other hand is precisely the weight lattice of $SU(n)/\mathbb Z_n$.  In the general simple case  $\Lambda^*(H)$ resulting from the Dirac quantization condition is the weight lattice $\Lambda(H^*)$ of the GNO dual group $H^*$ whose weight lattice is the dual weight lattice of $H$ and whose roots are identified with the coroots of $H$ \cite{Goddard:1976qe}. In addition the center and the fundamental group of $H^*$ are isomorphic to respectively the fundamental group and the center of $H$. Note that for all practical purposes the root system of $H^*$ can be identified with the root system of $H$ where the long and short roots are interchanged. \\
\\
We shall not repeat the proof of the duality of the center and the fundamental group, but we will sketch the proof of the fact that the root lattice of $H^*$ is always contained in the magnetic weight lattice. Finally we sketch the generalization to any connected compact Lie group. \\
If $H$ is not simply-connected we have $H=\widetilde H/Z$ where $\widetilde H$ is the universal cover of $H$ and $Z\subset Z(\widetilde H)$ a subgroup in the center of $\widetilde H$. Since $\Lambda(H) \subset \Lambda(\widetilde H)$ with $Z=\Lambda(\widetilde H)/\Lambda(H)$ the Dirac quantization condition (\ref{diraccond}) applied on $H$ is less restrictive than the condition for $\widetilde H$. Moreover one can check \cite{Goddard:1976qe}:
\begin{equation}
\Lambda^*(H)/\Lambda^*(\widetilde H)= \Lambda(\widetilde H)/\Lambda(H).
\end{equation}
This implies that the coroot lattice $\Lambda^*(\widetilde H)$  of $H$ is always contained in the magnetic weight lattice $\Lambda^*(H)$ of $H$ and in particular that any coroot $\alpha^*= \alpha/\alpha^2$ with $\alpha$ a root $H$, is contained in $\Lambda^*(H)$.\\
Without much effort this property can be shown to hold for any compact, connected Lie group.   Any such group $H$ say of rank $r$ can be expressed as:
\begin{equation}
H= \frac{U(1)^s\times K}{Z},
\end{equation}
where $K$ is a semi-simple and simply connected Lie group of rank $r-s$. The CSA of $H$ is spanned by $\{H_a:a=1,\dots,r\}$ where $H_a$ with $a\leq s$ are the generators of the $U(1)$ subgroups and $\{H_b: s < b \leq r\}$ span the CSA of $K$.  Any weight of $H$ can be expressed as $\lambda = (\lambda_1, \lambda_2)$ where $\lambda_1$ is a weight of $U(1)^s$ and $\lambda_2$ is a weight of $K$. Finally one finds that a magnetic charge $G_0$ defined by 
\begin{equation}
G_0 = \frac{4\pi}{e}  \alpha_j^*\cdot H,
\end{equation} 
where $\alpha_j$ is any of the $r-s$ simple roots of $H$, satisfies the quantization condition.\\
\begin{table}[!hbtp]
\begin{center}
\begin{tabular}{cc}
\hline
\hline
$H$ & $H^*$ \\
\hline
$SU(nm)/\mathbb Z_m$      &  $SU(nm)/\mathbb Z_n$ \\
$Sp(2n)$     &  $SO(2n+1)$           \\
$Spin(2n+1)$ &  $Sp(2n)/\mathbb Z_2$ \\
$Spin(2n)$   &  $SO(2n)/\mathbb Z_2$ \\
$SO(2n)$      &  $SO(2n)$            \\
$G_2$        &  $G_2$                \\
$F_4$        &  $F_4$                \\
$E_6$        &  $E_6/\mathbb Z_3$    \\
$E_7$        &  $E_7/\mathbb Z_2$    \\
$E_8$        &  $E_8$                \\ 
\hline
\hline \\         
\end{tabular}
\caption{Langlands or GNO dual pairs for simple Lie groups.}
\label{table:dualpairs}
\end{center} 
\end{table}
\begin{table}[!hbtp]
\begin{center}
\begin{tabular}{cc}
\hline
\hline
$H$ & $H^*$ \\
\hline
$(U(1)\times SU(n))/\mathbb Z_n      $      & $(U(1)\times SU(n))/\mathbb Z_n$            \\ 
$ U(1)\times Sp(2n)$                        & $U(1)\times SO(2n+1)$                       \\
$(U(1)\times Spin(2n+1))/\mathbb Z_2$       & $(U(1)\times Sp(2n))/\mathbb Z_2$           \\
$(U(1)\times Spin(2n))\mathbb Z_2$          & $(U(1)\times SO(2n))/\mathbb Z_2$           \\  
\hline
\hline \\
\end{tabular}
\caption{Examples of Langlands or GNO dual pairs for some compact Lie groups.}
\label{table:dualpairs2}
\end{center} 
\end{table}
\\
In this section we have identified the magnetic charge lattice of singular monopoles with the weight lattice of the dual group $H^*$ of the gauge group $H$. In table \ref{table:dualpairs} and \ref{table:dualpairs2} some examples are given of GNO dual pairs of Lie groups. Table \ref{table:dualpairs} is complete up to some dual pairs related to $Spin(4n)$ that are obtained by modding out non-diagonal $Z_2$ subgroups of the center $\mathbb Z_2\times \mathbb Z_2$. The GNO dual groups for these cases can be found in \cite{Goddard:1976qe}. In section \ref{subsect:quantBPS} we shall briefly explain how the dual pairing in table \ref{table:dualpairs2} is determined. \\
\\
The magnetic charge lattice contains an important subset which we shall need later on: even if one restricts $G_0$ to the CSA there is some gauge freedom left which corresponds to the action of the Weyl group. Modding out this Weyl action gives a set of equivalence classes of magnetic charges which are naturally labelled by dominant integral weights in the weight lattice of $H^*$.
\subsection{Quantization condition for smooth monopoles}
\label{subsect:quantsmooth}
Yang-Mills-Higgs theories have solutions that behave at spatial infinity as singular Dirac monopoles but which are nonetheless completely smooth at the origin. This is possible if one starts out with a compact, connected, semi-simple gauge group $G$ which is spontaneously broken to a subgroup $H$. Since all the fields are smooth, the gauge field defines a connection of a principal $G$-bundle over space which we take to be $\mathbb R^3$. The Higgs field 
is a  section of a the adjoint bundle. As $\mathbb R^3$ is contractible 
the principal $G$-bundle is automatically trivial, so $\Phi$ is simply
a Lie-algebra valued function.
We would like to impose boundary conditions for  the Higgs field  $\Phi$ and the magnetic field $B$ at spatial infinity
which ensure that the total energy carried by a  solution of the Yang-Mills-Higgs equations is finite. To our
knowledge the question of 
which conditions are necessary and sufficient 
 has not  been answered in general. Below we review
some standard arguments, many of them summarized in \cite{GoddardOlive1978}.\\
\\ 
We assume an energy functional for static fields of the usual form
\begin{equation}
\label{energy}
E[\Phi,A]=\int\, \frac 1 2 |D_k\Phi|^2+\frac 1 2 |B_k|^2 +V(\Phi)\, d^3 x ,
\end{equation}
where $D_k=\partial_k-ieA_k$ is the covariant derivative with respect 
to the $G$-connection $A$, and the magnetic field 
is given by  $-ieB_k=- \frac 1 2 ie \epsilon_{klm}F_{lm} = \frac 1 2 \epsilon_{klm}[D_l,D_m]$.
The potential $V$ is a $G$-invariant function on the Lie algebra of $G$ 
whose minimum is attained for 
 non-vanishing value of $|\Phi|$; the set of minima is called the 
vacuum manifold. The variational equations for 
this functional are 
\begin{equation}
\label{fieldeq}
\epsilon_{klm}D_lB_m= ie[\Phi, D_k\Phi], \qquad D_kD_k\Phi=
\frac{\partial V}{\partial \Phi}.
\end{equation}
In order to ensure that solutions of these equations have finite energy
we  require the fields $\Phi$ and $B_i$ to have the following asymptotic form
for large $r$:
\begin{equation}
\label{eqn:boundcond}
\begin{split}
\Phi &=  \phi(\hat r) + \frac{ f(\hat r)}{4\pi r} + \mathcal O\left(r^{-(1+\delta)}\right) \qquad r \gg 1 \\
B    &= \; \frac{ G(\hat r)}{4\pi r^2}\hat r \; + \;\mathcal O\left(r^{-(2+\delta)}\right) \qquad\qquad\!\! r \gg 1.
\end{split}
\end{equation}   
Here $\delta > 0$ is some constant and $\phi(\hat r),  f(\hat r),$ and $G(\hat r)$ are smooth functions  on $S^2$ taking values in the Lie algebra of the gauge group $G$ which have to 
 satisfy various  conditions.\\
First of all, the function $\phi$ has to take values in the  vacuum manifold of the potential $V$. It is thus  a smooth map from the two-sphere 
to that vacuum manifold. The homotopy class of that map defines the monopole's topological charge \cite{GoddardOlive1978}. Since the vacuum manifold can be identified with the coset space $G/H$ the topological charge takes value in $\pi_2(G/H)$. Secondly, writing $\nabla$ for the induced exterior covariant derivative tangent to the two-sphere ``at infinity'' it is easy to check that 
\begin{equation}
\label{phifcond}
 \nabla\phi=0,\qquad  \nabla f = 0
\end{equation}
are necessary conditions for the integral defining the energy \eqref{energy} to converge. The first of these equations  implies
\begin{equation}
\label{nocom}
[\phi(\hat r),G(\hat r)]=0.
\end{equation}
The quickest way to to see this is to note that the curvature on the two-sphere at infinity is
\begin{equation}
F^\infty=
*(\frac{G(\hat r)}{4\pi r^2} dr)=\frac{G(\hat r)}{4 \pi}\sin\theta d\theta \wedge d\varphi.
\end{equation}
Since $[\nabla, \nabla]=-ieF^\infty$, it follows that  $\nabla \phi=0$
implies $[F^\infty,\phi]=0$.
Finally we also  require that 
\begin{equation}
\label{Gcond}
\nabla G=0.
\end{equation}
and that 
\begin{equation}
\label{fcond}
[\phi(\hat r), f(\hat r)]=0.
\end{equation}
The condition \eqref{Gcond} is crucial for what follows, and seems to be satisfied for all known finite energy solutions 
\cite{GoddardOlive1978}. The  condition \eqref{fcond} is required so that  the first of the equations \eqref{fieldeq} is satisfied
 to lowest order when  the expansion \eqref{eqn:boundcond} is inserted. In general there will be additional requirements on the functions $\phi$ and 
$f$ that depend on the precise form of the potential $V$ in \eqref{energy}. Since we do not specify $V$ we will not discuss these further.\\
\\
The above conditions can be much simplified by changing gauge. The equations \eqref{phifcond} and \eqref{Gcond} imply  that for each of 
the Lie-algebra valued functions $\phi,f$ and $G$  the values at any two points 
on the two-sphere at infinity  are conjugate to one another (the 
required conjugating element being the parallel transport along the path
connecting the points). We can  therefore
pick a point $\hat r_0$, say the north pole, and gauge transform
$\phi$ into $\Phi_0=\phi(\hat r_0)$, $f$ into $\Phi_1=f(\hat r_0)$ and 
$G$ into $G_0=G(\hat r_0)$. However, since $S^2$ is not contractible,
we will, in general,  not be able to do this smoothly everywhere
on the two-sphere at infinity. 
If, instead, we cover the two-sphere with two contractible patches
which overlap on the equator, then there are smooth  
 gauge transformations 
$g_+$ and $g_-$ defined, respectively, on the northern and southern hemisphere,
so that the following equations  hold where they are  defined:
\begin{align}
\label{gaugetra}
\phi(\hat r )&=  g_\pm^{-1}(\hat r )\Phi_0 g_\pm(\hat r) \\
f(\hat r )&=  g_\pm^{-1}(\hat r )\Phi_1 g_\pm(\hat r) \\
G(\hat r )&= g_\pm^{-1}(\hat r )G_0 g_\pm(\hat r).
\end{align}
After applying these gauge transformation, our bundle is defined in two patches, with transition function $\mathcal G = g_+g_-^{-1}$ defined near the equator. This transition function leaves $\Phi_0$ invariant, and hence lies in the subgroup $H$  of $G$ which stabilizes $\Phi_0$. This, by definition, is the residual or unbroken gauge group referred to in the opening paragraph of this section. It follows from \eqref{nocom}, that $[\Phi_0,G_0]=0$, so that 
$G_0$ lies in the Lie algebra of $H$. Similarly, \eqref{fcond} implies that $\Phi_1$ lies in the Lie algebra of $H$.
After applying the local gauge transformations
\eqref{gaugetra}, the asymptotic form of the fields is 
\begin{equation}
\label{eqn:boundsmooth}
\begin{split}
\Phi &= \Phi_0 + \frac{\Phi_1}{4\pi r}\!\, + \!\,\mathcal O \!\left(r^{-(1+\delta)}\right) \\
B &= \,\frac{G_0}{4\pi r^2} \hat r \,\, +\,\, \mathcal O\left(r^{-(2+\delta)}\right). 
\end{split}
\end{equation}
Note that ``the Higgs field at infinity'' is now constant, taking the value $\Phi_0$ everywhere. In particular, it therefore belongs to the trivial homotopy class of maps from the two-sphere to the vacuum manifold. The topological charges originally encoded in the map $\phi$ can no longer be computed from the Higgs field. Instead they are now encoded in transition function $\mathcal G$. Since, in the new gauge, the magnetic field at large $r$ is that of a  Dirac monopole with gauge group $H$ we can relate the transition function to the magnetic charge as before:
\begin{equation}
\label{eqn:equator2}
\mathcal G(\varphi)= \exp\left(\frac{ie}{2\pi}G_0\varphi\right)
\end{equation} 
We thus obtain a quantization condition for the magnetic charge of smooth monopoles, following the same arguments as in the singular case. For each representation of $H$ the gauge transformation must be single valued if one goes around the equator, so that 
\begin{equation}
\label{diraccond2}
2\lambda \cdot g \in \mathbb Z,  
\end{equation}
for all charges $\lambda$ in the weight lattice of $H$.\\
\\
One observes that the magnetic charge lattice of smooth monopoles lies in the weight lattice of the GNO dual group $H^*$. There is, however, another consistency condition \cite{Englert:1976ng}. Note that a single valued gauge transformation on the equator defines a closed curve in $H$ as well as in $G$, starting and ending at the unit element. Since the original  $G$-bundle is trivial,  this closed curve has to be contractible in $G$. Therefore the monopole's topological charge is labelled by an element in $\pi_1(H)$ which maps to a trivial element in $\pi_1(G)$.  This is consistent with our earlier remark that the topological charge is an element of  $\pi_2(G/H)$ because of the isomorphism $\pi_2(G/H)\simeq \text{ker}(\pi_1(H)\rightarrow \pi_1(G))$.\\
\\
To find the appropriate charge lattice we use the fact that a loop in $G$ is trivial if and only if its lift to the universal covering group $\widetilde G$ is also a loop (closed path).  This implies that for smooth monopoles the quantization condition should not be evaluated in the group $H$ itself but instead in the group $\widetilde H \subset \widetilde G$ defined by the Higgs VEV $\Phi_0$. Consequently equation (\ref{diraccond2}) must not only hold for all representations of $H$ but in fact for all representations of $\widetilde H$. Note that if $G$ is simply connected then $\widetilde H = H$. In the next section we shall work this topological condition out in more detail.
\subsection{Quantization condition for smooth BPS monopoles}
\label{subsect:quantBPS}
In this paper we will mainly focus on BPS monopoles in spontaneously broken theories. We shall therefore work out some results of the previous section in somewhat more detail for the BPS case.  We shall also give an explicit description of the magnetic charge lattice. In addition we introduce terminology that is conveniently used in the remainder of this paper.\\
\\
By BPS monopoles we mean static, finite energy solutions of the BPS equations
\begin{equation}
\label{bpseq}
B_i= D_i \Phi
\end{equation}
 in a Yang-Mills-Higgs theory with a compact, connected,  semi-simple gauge group $G$. The equations \eqref{bpseq} imply the second order
equations \eqref{fieldeq}. In order to obtain finite energy solutions
we again impose the boundary conditions \eqref{eqn:boundcond}. As in the 
previous section we can gauge transform these into the form 
\eqref{eqn:boundsmooth}.
 There are some differences with the non-BPS case. The potential $V$ in \eqref{energy} vanishes in the BPS limit, so does not furnish any conditions on the functions $\phi$ and $f$. On the other hand, by substituting (\ref{eqn:boundsmooth}) in the BPS equation and solving order by order one finds that $f=-G$, or, equivalently, $\Phi_1=-G_0$. As before we have $[\Phi_0,G_0]=0$, so in the BPS case we automatically  have $[\Phi_0,\Phi_1]=0$.
From now on we shall thus define a BPS monopole to be a smooth solution of the BPS equations satisfying the boundary condition \eqref{eqn:boundcond} with  $\Phi_1=-G_0$. After applying  the local gauge transformations discussed in 
the previous section, these boundary conditions are equivalent to
\begin{equation}
\label{eqn:BPSbc}
\begin{split}
\Phi &= \Phi_0 - \frac{G_0}{4\pi r} + \mathcal O \left(r^{-(1+\delta)}\right) \\ 
B &=\, \frac{G_0}{4\pi r^2} \hat r\: + \:\mathcal O\left(r^{-(2+\delta)}\right),
\end{split}
\end{equation}
where $\Phi_0$ and $G_0$ are commuting  elements in the Lie algebra of $G$. These boundary conditions are sufficient to guarantee that the energy of the BPS monopole is finite. It is in general not known what the necessary boundary conditions are to obtain a finite energy configuration. It is expected though \cite{Jarvis1,Murray:2003mm}, and true for $G= SU(2)$ \cite{Jaffe}, 
that the boundary conditions above follow from the finite energy condition and the BPS equation. \\
\\
Before we give an explicit description of the magnetic charge lattice let us summarize some properties of the residual gauge group. Since $[\Phi_0, G_0]=0$ there is a gauge transformation that maps $\Phi_0$ and $G_0$ to our chosen CSA of $G$. Without loss of generality we can thus express $\Phi_0$ and $G_0$ in terms of the generators $\left( H_a \right)$ of that CSA:
\begin{equation}
\begin{split}
\Phi_0&= ~~ \mu\cdot H \\
G_0&= \frac{4\pi}{e}g\cdot H. 
\end{split}
\end{equation}\\
The residual gauge group is generated by generators $L$ in the Lie algebra of $G$ satisfying $[L,\Phi_0]=0$. Since generators in the CSA by definition commute with the Higgs VEV the residual group $H$ contains at least the maximal torus $U(1)^r\subset G$.  For generic values of the Higgs VEV this is the complete residual gauge symmetry. If the Higgs VEV is perpendicular to a root $\alpha$ the residual gauge group becomes non-abelian. This follows from the action of the corresponding ladder operator  $E_\alpha$ in the Cartan-Weyl basis on the Higgs VEV:  $[E_\alpha,\Phi_0]=- \mu\cdot\alpha\, E_\alpha=0$. Accordingly we shall call a root of $G$ \textit{broken} if it has a non-vanishing inner product with $\mu$ and we shall define it to be \textit{unbroken} if this inner product vanishes.\\
The residual gauge group is locally of the form $U(1)^s\times K$, where $K$ is some semi-simple Lie group. The root system of $K$ is derived from the root system of $G$ by removing the broken roots. Similarly, the Dynkin diagram of $K$ is found from the Dynkin diagram of $G$ by removing the nodes related to broken simple roots. For completeness we finally define a fundamental weight to be (un)broken if the corresponding simple root is (un)broken.\\
\\
The magnetic charge lattice for smooth monopoles lies in the dual weight lattice of $H$, as we saw  in the previous chapter. For adjoint symmetry breaking the weight lattice of $H$ is isomorphic to the weight lattice of $G$. Moreover the isomorphism respects the action of the Weyl group $W(H)\subset W(G)$. The existence of an  isomorphism between $\Lambda(G)$ and $\Lambda(H)$ is easily understood since the weight lattices of $H$ and $G$ are determined by the irreducible representations of their maximal tori which are isomorphic for adjoint symmetry breaking. A natural choice for the CSA of $H$ is to identify it with the CSA of $G$. In this case $\Lambda(G)$ and $\Lambda(H)$ are not just isomorphic but also isometric. Since the roots of $H$ can be identified with roots of $G$ and since the Weyl group is generated by the reflections in the hyperplanes orthogonal to the roots, this isometry obviously respects the action of $W(H)$.  Often the CSA of $H$ is identified with the CSA of $G$ only up to normalization factors. This leads to rescalings of the weight lattice of $H$. Of course one can apply an overall rescaling without spoiling the invariance of weight lattice under the Weyl reflections.  One can also choose the generators of  $U(1)^s$-factor such that the corresponding charges are either integral or half-integral. Note that these rescalings again respect the action of $W(H)$. To avoid confusion we shall ignore these possible rescalings in the remainder of this paper and take $\Lambda(H)$ to be isometric to $\Lambda(G)$.\\
Since the weight lattices $\Lambda(H)$ and $\Lambda(G)$ are isometric their dual lattices $\Lambda^*(H)$ and $\Lambda^*(G)$ are isometric too. We thus see that we the Dirac quantization condition (\ref{diraccond2}) for adjoint symmetry breaking can consistently be evaluated in terms of either $H$ or $G$.\\
\\
Remember that for smooth monopoles monopoles there is yet another condition:  since one starts out from a trivial $G$ bundle the magnetic charge should define a topologically trivial loop in $G$ as explained in the previous section. For general symmetry breaking this implies that the Dirac quantization condition must be evaluated with respect to weight lattice of $\widetilde H \subset \widetilde G$, where $\widetilde G$ is the universal covering group of $G$. For adjoint symmetry breaking we can consistently lift the quantization condition to $G$; the weight lattice of $\widetilde H$ is isometric to the weight lattice of $\widetilde G$. The weight lattice of $\widetilde G$ is generated by the fundamental weights $\{\lambda_i\}$ and hence the magnetic charge lattice for smooth BPS monopoles is given by the solutions of:      
\begin{equation}
\label{diraccond3}
2\lambda_i \cdot g \in \mathbb Z,  
\end{equation}
for all fundamental weights $\lambda_i$ of $\widetilde G$. The most general solution of this equation is easily solved in terms of the simple coroots of $G$: 
\begin{equation}
\label{eqn:corootcon}
g= \sum_i m_i \alpha_i^* \qquad m_i \in \mathbb Z,
\end{equation}
with $\alpha_i^*=\alpha_i/\alpha_i^2$ and $\{\alpha_i\}$ the simple roots of $G$.\\
We thus conclude that the magnetic charge lattice for smooth BPS monopoles is generated by the simple coroots of $G$. The resulting coroot lattice $\Lambda^*(\tilde G)$ corresponds precisely to the weight lattice $\Lambda({\widetilde G}^*)$  of the GNO dual group ${\widetilde G}^*$ as mentioned in section \ref{sect:singcondition}. Similarly, the dual lattice $\Lambda^*(\widetilde H)$ can be identified with $\Lambda({\widetilde H}^*)$. With $\Lambda^*(\widetilde G)$ being isometric to $\Lambda^*(\widetilde H)$ we now conclude that the weight lattice of ${\widetilde G}^*$ can be identified with the weight lattice of ${\widetilde H}^*$. For $G$ simply connected we have thus established an isometry between the root lattice of $G^*$ and the weight lattice of $H^*$. We have used this isometry to compute the GNO dual pairs given in table \ref{table:dualpairs2} which  appear in the minimal adjoint symmetry breaking of the classical Lie groups.\\
\\
Above we have seen that the magnetic charge lattice for smooth BPS mono\-poles corresponds to the coroot lattice of the gauge group $G$. One can split the set of coroots into broken coroots and unbroken coroots.  A coroot is defined to be broken or unbroken if the  corresponding root is respectively broken or unbroken. Note that the unbroken coroots are precisely the roots of $H^*$. The distinction between broken and unbroken applies in particular to simple coroots. There is however alternative terminology for the components of the magnetic charges that reflects these same properties. Broken simple coroots are identified with \emph{topological charges} while unbroken simple coroots are related to so-called \emph{holomorphic charges}.\\
Remember that the magnetic charge  charge $g=m_i\alpha_i^*$ defines an element in $\ker(\pi_1(H) \to \pi_1(G))$.  One might hope that every single magnetic charge $g$, i.e.~every point in the coroot lattice, defines a unique topological charge. If in that case a static monopole solution does indeed exist even its stability under smooth deformations is guaranteed.  Such a picture does hold for maximally broken theories where the residual gauge group equals the maximal torus $U(1)^r\subset G$. If $H$ contains a non-abelian factor the situation is slightly more complicated because these factors are not detected by the fundamental group. For $G$ equal to  $SU(3)$  for instance the magnetic charge lattice is 2-dimensional and $\pi_1\left(SU(3)\right)=0$. In the maximally broken theory we have $\pi_1(U(1)\times U(1))=\mathbb Z\times \mathbb Z$, while for minimal symmetry breaking $\pi_1(U(2))=\pi_1(U(1))=\mathbb Z$. As a rule of thumb one can say that the components of the magnetic charges related to the $U(1)$-factors in $H$  are topological charges. It should be clear that these components correspond to the broken simple coroots. We therefore call the coefficients $m_i = 2\lambda_i \cdot g$ with $\lambda_i$ a broken fundamental weight the topological charges of $g$. The remaining components of $g$ are often called holomorphic charges.
\subsection{Murray condition}
\label{subsect:murraycond}
\begin{figure}[!hb]
\centerline{
\includegraphics{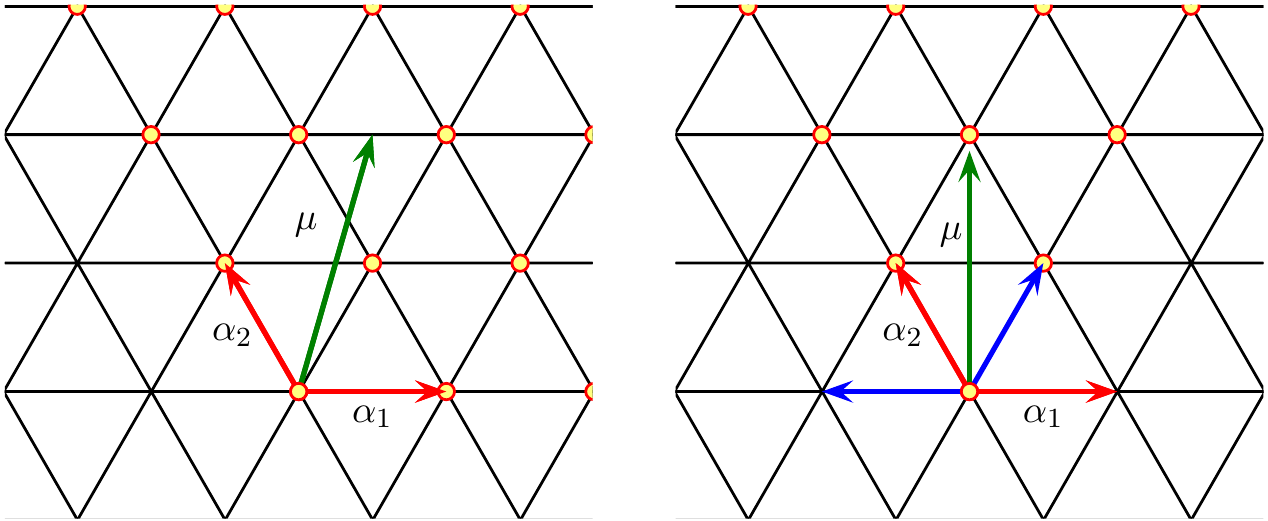}
}
\caption{The Murray cone for $SU(3)$ as a subset of the Cartan subalgebra. If the residual gauge group equals $U(1)\times U(1)$ (left) the Higgs VEV determines a unique set of simple roots. The static BPS monopoles have magnetic charges equal to a positive linear combination of these roots. These charges are in one-to-one correspondence with the positive topological charges. If the residual gauge group is $U(2)$ (right) there are two choices of simple roots. Only those charges that have a positive expansion for both these choices correspond to non-empty moduli spaces of static BPS monopoles. There is only a single topological charge which is proportional to the inner product of the magnetic charge with the Higgs VEV $\mu$. As can been seen from the picture the total magnetic charge is not uniquely determined by the topological charge alone: non-abelian monopoles may carry non-trivial holomorphic charges.}
\label{fig:conesu3}
\end{figure}

We have found that magnetic charges of smooth monopoles in a Yang-Mills-Higgs theory lie on the coroot lattice of the gauge group. In the BPS limit there is yet another consistency condition which was first discovered by Murray for $SU(n)$ \cite{Murray:1989zk}. We refer to this condition as \emph{the Murray condition} even though its final formulation for general gauge groups stems from a paper by Murray and Singer \cite{Murray:2003mm}. For a derivation of the Murray condition we refer to these original papers. We shall only briefly review some properties of roots  which are crucial for the Murray condition. Next we shall formulate the results of Murray and Singer in such a way that the set of magnetic charges for BPS monopoles can easily be identified. Finally we show that our formulation is equivalent to the  condition as stated in \cite{Murray:2003mm}. Both formulations of the Murray condition will show up in later sections. The set of magnetic charges satisfying the Murray condition shall be called the Murray cone. At the end of this section we shall also introduce the fundamental Murray cone.\\
\\
The Murray condition hinges on the fact that one can split the root system of $G$ into positive and negative roots with respect to the Higgs VEV. If for a root $\alpha$ we have $\alpha\cdot\mu > 0$ it is by definition positive and if $\alpha\cdot\mu< 0$ it is negative. The set of roots is now partitioned into two mutually exclusive sets, at least if the residual gauge group is abelian. In that case we can as usual define a simple root to be a positive root that cannot be expressed as a sum of two other positive roots and it turns out that the Higgs VEV defines a unique set of simple roots. These form a basis of the root diagram is such a way that every positive root is a linear combination of simple roots with positive coefficients and similarly every negative root is a linear combination with negative coefficients. In the non-abelian case there exist roots such that $\alpha\cdot\mu=0$.  Hence there are several choices for a set of simple roots which are consistent with the Higgs VEV. Again for a fixed choice such simple roots must by definition have the property that all roots are a linear combination of simple roots with either only positive or only negative coefficients. In addition the simple roots must have either a strictly positive or a vanishing inner product with the Higgs VEV:
\begin{equation}
\label{eqn:closure}
\alpha_i\cdot \mu \geq 0.
\end{equation}
This condition implies that $\mu$ must lie in the closure of the fundamental Weyl chamber. In the remainder of this paper we shall always choose simple roots so that the inequality in (\ref{eqn:closure}) is satisfied.\\
All choices for a set of simple roots respecting the Higgs VEV are related by the residual Weyl group $W(H)$. This is seen as follows. In general all choices of simple roots in the root system of $G$ are related by the Weyl group $W(G)$ of $G$. Since Weyl transformations are orthogonal we have for all $w\in W(G)$ $w(\alpha_i)\cdot\mu= \alpha_i\cdot w^{-1}(\mu)$. Given a set of positive roots satisfying (\ref{eqn:closure})  the action of $w \in W(G)$  gives another set of simple roots satisfying the same condition if and only if $\mu$ and $w(\mu)$ lie in the closure of same Weyl chamber. This is only possible if $\mu$ is actually invariant under $w$, implying that $w\in W(H)\subset W(G)$.\\  
\\
Above we have defined a positivity condition for the roots of $G$ that is consistent with the Higgs VEV. This same definition is applicable for coroots since these differ from the roots by a scaling. We now also extend this definition of positivity in a consistent way to the complete (co)root lattice. We call an element on the (co)root lattice positive if it is a linear combination of simple (co)roots with positive integer coefficients. Note that the intersection of the set of positive elements in the (co)root lattice with the set of (co)roots is precisely the set of positive (co)roots. Finally we see that if the Higgs VEV lies in the fundamental Weyl chamber then the innerproduct of any positive element in the (co)root lattice with $\mu$ is non-negative.\\  
\\
Murray and Singer have found  that the magnetic charge must be positive with respect to all possible choices of simple roots consistent with the Higgs VEV. This means that in the expansion $g=\sum_i m_i\alpha^*_i$ the coefficients $m_i$ should be positive for all possible choices of simple roots $\left(\alpha_i\right)$ that satisfy $\alpha_i\cdot \mu \geq 0$. 
The Murray condition can be summarized as follows:
\begin{equation}
\label{murraycond}
2w(\lambda_i)\cdot g \geq 0 ~~ \forall w\in W(H),~\forall \lambda_i.
\end{equation} 
This is seen from the fact that the fundamental weights and simple roots satisfy $2\lambda_i\cdot\alpha^*_j=\delta_{ij}$ and that all allowed choices of positive simple roots and fundamental weights are related by the residual Weyl group $W(H)\subset W(G)$.\\
\\
The Murray condition defines a solid cone in the CSA. In combination with the Dirac quantization condition this results in a discrete cone of magnetic charges. We shall call this cone the Murray cone. As an example one can consider $SU(3)$ broken to either $U(1)\times U(1)$ or $U(2)$ as depicted in figure \ref{fig:conesu3}. In the first case the Weyl group of the residual gauge group is trivial and the Murray condition simply implies that the topological charges must be positive. In the second case the residual Weyl group is $\mathbb Z_2$, the reflections in the line perpendicular to $\alpha_1$. Consequently there are two possible choices of positive simple roots which makes the Murray condition more restrictive. The topological charge still has to be positive, just like the holomorphic charge, but the holomorphic charge is bounded by the topological charge. \\
\\
We shall finish this section with yet another formulation of the Murray condition originating from proposition 4.1 in the paper of Murray and Singer \cite{Murray:2003mm}. It relies on the fact that the holomorphic charges can be minimized under the action of the residual Weyl group. For any element $g$  in the coroot lattice there exists a uniquely determined reduced magnetic charge $\tilde g$ in the Weyl orbit of $g$ such that $\alpha_j\cdot \tilde g \leq 0$  for all unbroken simple roots $\alpha_j$. The Murray condition can be expressed in terms of this minimized charges. A magnetic charge $g$ is positive with respect to any chosen set of simple roots if and only if for a fixed choice of simple roots  its reduced magnetic charge is positive. The reduced magnetic charge should thus satisfy:
\begin{equation}
\label{murraycond3}
2\lambda_i\cdot \tilde g \geq 0 ~~\forall \lambda_i.
\end{equation} 
We shall shortly show that $\tilde g$ does indeed exist and is unique. But already we can see that this last condition easily follows from (\ref{murraycond}). Since $\tilde g=\tilde w(g)$ for some $\tilde w\in W(H)$ we have $w(\lambda_i)\cdot \tilde g = w(\lambda_i)\cdot \tilde w(g) = \tilde w^{-1}\left(w(\lambda_i)\right)\cdot g= w'(\lambda_i)\cdot g \geq 0$, where $w'=\tilde w^{-1}w\in W(H)$.   To show equivalence however we also have to show that (\ref{murraycond}) follows from (\ref{murraycond3}), which boils down to proving the following proposition:
\begin{prop}
\label{prop:redhol}
If the reduced magnetic charge $\tilde g$ is positive then $w(\tilde g)$ is positive for all $w\in W(H)$.
\end{prop}
\proof We take the gauge group $G$ broken to $H$. The magnetic charges of BPS monopoles lie on the coroot lattice of $G$ or equivalently the root lattice of $G^*$. We can  assume $G$ to be simply-connected since this does not affect the magnetic charge lattice. Under this assumption there is an isomorphism $\lambda$ from  the coroot lattice $\Lambda^*(G)$ to  the weight lattice $\Lambda(H^*)$ of $H^*$ as discussed in section \ref{subsect:quantBPS}. Up to discrete factors $H^*$ is of the form $U(1)^s\times K^*$, where $K^*$ is some semi-simple Lie group. Similarly, the set of simple roots of $G$ is split up into $s$ broken roots $\{\alpha_i\}$ with $0< i \leq s$ and $r-s$ unbroken roots $\{\alpha_j\}$ with $ s <j \leq r$. The magnetic charges are thus expanded as $g= \sum_i m_i\alpha^*_i + \sum_j h_j \alpha^*_j$.\\
The linear map $\lambda$ is defined by the images of the simple coroots. For the unbroken simple coroots this is particularly simple. We have $\lambda(\alpha^*_j)=\alpha^*_j$. More generally the image is given in terms of the abelian charges and a weight of $K^*$. While the abelian charges are identified with the topological charges $\{m_i\}$ the non-abelian charge can be expanded in terms of the fundamental weights $\lambda_j$ of $K^*$. The  coefficients, i.e.~the Dynkin labels, are given by the projection on the roots of $K^*$: $k_j= 2\alpha^*_j\cdot g/{\alpha_j^*}^2$. Being sums of multiples of the entries of the Cartan matrix of $G^*$ these labels are indeed integers.\\
We can now easily prove that the reduced magnetic charge $\tilde g$ exists and is unique. Let $h:= \lambda(g)$. Any weight $h\in \Lambda(H^*)$ can be mapped to a unique weight $\tilde h$ in the anti-fundamental Weyl chamber via a Weyl transformation. We thus have $\tilde h \cdot \alpha \leq 0$. The reduced magnetic charge $\tilde g$ is fixed by $\lambda(\tilde g)=\tilde h$. Since $2\lambda(g)\cdot \alpha^*_j/{\alpha^*_j}^2 =  2g\cdot \alpha^*_j/{\alpha^*_j}^2$ we have  $\alpha^*_j\cdot g\leq 0$ for all unbroken roots of $G^*$. The same inequality holds for the unbroken roots of $G$ itself.\\
We now return to the proof of the proposition. First we shall use the fact that $\lambda$ respects the residual Weyl group is the sense that $\lambda\left(w(g)\right)=w\left(\lambda(g)\right)$ for all $w\in W(H)$. This can be proved using the fact that any Weyl transformation is a sequence of Weyl reflections $w_j$ in the hyperplanes perpendicular to the  simple coroots $\alpha^*_j$. It is thus sufficient to prove that $\lambda$ commutes with $w_j$ for all unbroken simple roots. We have
\begin{equation}
\begin{split}
\lambda\left(w_j(g)\right)&= \lambda\left( g - \frac{2g\cdot\alpha^*_j}{{\alpha^*_j}^2}\alpha^*_j\right) 
                                       \:\,= \lambda(g) -\frac{2g\cdot\alpha^*_j}{{\alpha^*_j}^2}\lambda(\alpha^*_j) \\
                          &= \lambda(g) - \frac{2\lambda(g)\cdot\alpha^*_j}{{\alpha^*_j}^2}\alpha^*_j 
                             = w_j\left(\lambda(g)\right).
\end{split}
\end{equation}
Note that for the unbroken roots $\lambda(\alpha_j)=\alpha_j$ and that $\lambda$ is an isometry as discussed in section \ref{subsect:quantBPS} and thus leaves the innerproduct invariant.\\
Secondly for the proof of the proposition we use the fact that for a lowest weight $\tilde h$ we have $w(\tilde h)= \tilde h + n_j\alpha^*_j$ with $n_j\geq 0$ for any $w\in W(H^*)$,  see for example chapter 10 to 13 of \cite{Humphreys:1980dw}. For $\tilde g$ and any $w\in W(H^*)=W(H)$ we now get:
\begin{equation}
\begin{split}
\lambda\left(w(\tilde g)\right) &= w\left( \lambda(\tilde g)\right) \;= w( \tilde h )\\
                                &= \tilde h + n_j\alpha^*_j = \lambda(\tilde g) + n_j\lambda(\alpha^*_j)\\
                                &= \lambda( \tilde g + n_j\alpha^*_j). 
\end{split}
\end{equation}
Consequently in terms of the unbroken simple coroots of $G$ we find  $w(\tilde g) = \tilde g + n_j\alpha^*_j$ where $n_j\geq 0$. Thus for the all fundamental weights of $G$ we have $2\lambda_i\cdot w(\tilde g) \geq 0$ if  $2\lambda_i \cdot \tilde g \geq 0$.\endproof 
Note that the set of positive reduced magnetic charges is a subset of the Murray cone and can be obtained  by modding out the residual Weyl group. 
The set of Weyl orbits in the Murray cone is a physically important object; it corresponds to the magnetic charge sectors of the theory. This follows from the fact that a magnetic charge $g$ is defined only modulo the action of the residual Weyl group. 
For this reason we shall introduce a set called the fundamental Murray cone which is bijective to the set of of Weyl orbits in the Murray cone. The set of positive reduced magnetic charges can of course be identified with the fundamental Murray cone. However, it would be more appropriate  to call this set the anti-fundamental Murray cone. We recall that a reduced magnetic charge $\tilde g$ satisfies $\alpha_j\cdot \tilde g \leq 0$ for all unbroken simple roots $\alpha_j$. It follows from this condition that $\tilde g$ can be identified with a lowest weight of $H^*$. Similarly, we can define the subset of the Murray cone $\{g:\alpha_j\cdot \tilde g \geq 0\}$. These magnetic charges now map to the fundamental Weyl chamber of $H^*$, hence we call this set the fundamental Murray cone. We thus find that the magnetic charge sectors are labelled by dominant integral weights of the residual gauge group. A similar conclusions was drawn for singular monopoles by Kapustin \cite{Kapustin:2005py}.\\ 
\section{Generating charges}
\label{sect:gencharges}
As we have seen in the last section consistency conditions on the charges of magnetic monopoles give rise to certain discrete sets of magnetic charges. In the case of singular monopoles this set is nothing but the weight lattice of the dual group $H^*$.  The set of charges of smooth monopoles in a theory with adjoint symmetry breaking corresponds to the root lattice of the dual group $G^*$. Alternatively one can view this set as a subset in the weight lattice of the residual dual gauge group $H^*\subset G^*$. In the BPS limit the minimal energy configurations satisfy an even stronger condition which gives rise to the so-called Murray cone in the root lattice of $G^*$. Both the weight lattice of $H^*$ and the Murray cone in the root lattice of $G^*$ contain an important subset which is obtained by modding out the Weyl group of $H^*$. For singular monopoles one simply obtains the set of dominant integral weights, i.e.~the fundamental Weyl chamber of $H^*$. In the case of smooth BPS monopoles modding out the residual Weyl group is equivalent to restricting the charges to the fundamental Murray cone.\\
\\
In each case we want to find a set of minimal charges that generate all remaining charges via positive integer linear combinations. As it turns out this problem is most easily solved for the Murray cone. In the latter case the generators  can be identified as the coroots with minimal topological charges. Below we shall prove this for any compact, connected semi-simple Lie group $G$ and arbitrary symmetry breaking.  For the weight lattice $\Lambda(H^*)$ one can give a generic description for a small set of generators. To find a smallest set of generators one needs to know some detailed properties of $H^*$.  The generators the fundamental Weyl chamber and the fundamental Murray cone are not easily identified in general either. In all these cases we shall therefore restrict ourselves to some clear examples.\\
\\
The physical interpretation of the generating charges is that the monopoles with these minimal charges are the building blocks of all monopoles in the theory. We shall therefore call monopoles with minimal charges in the weight lattice of $H^*$ or in the Murray cone in $G^*$ \textit{fundamental monopoles}. The monopoles corresponding to the generators of the fundamental Weyl chamber and those related to the fundamental Murray cone both are called \textit{basic monopoles}. In section \ref{sect:modspace} and \ref{sect:fusionprop} we study to what extent these notions make sense in the classical theory.
\subsection{Generators of the Murray cone}
\label{subsect:genmurray}
Given two allowed magnetic charges $g$ and $g'$, that is two magnetic charges satisfying the Dirac condition (\ref{diraccond}) and the Murray condition (\ref{murraycond}), one can easily show that the linear combination $ng+ n'g'$ with $n,n'\in \mathbb N$ again is an allowed magnetic charge. This raises the question whether all allowed magnetic charges can be generated from a certain minimal set of charges. This would mean that all charges can be decomposed as linear combinations of these generating charges with positive integer coefficients. The minimal set of generating charges is precisely the set of indecomposable charges. These indecomposable charges cannot be expressed as a non-trivial positive linear combination of charges in the Murray cone. It is obvious that such a set exists. It is also not difficult to show that such a set is unique. This follows from the fact all negative magnetic charges are excluded by the Murray condition. Despite its existence and uniqueness we do not know a priori what the set of generating charges is, let alone that we can be sure it is reasonably small or even finite. \\
\\
There are some charges which are certainly part of the generating set, namely those for which the corresponding topological charges are minimal. These are the allowed charges $g$ such that $2\lambda_i\cdot g = 1$ for one particular broken fundamental weight $\lambda_i$ and $2\lambda_j\cdot g = 0$ for all other broken fundamental weights. 
\begin{prop}
Topologically minimal charges are indecomposable.
\end{prop}
\proof If an allowed charge $g$ with a minimal topological component can be decomposed into two allowed charges, $g=g'+g''$ then one of these, say $g'$, would have a topological component equal to zero. This means that $2\lambda_i\cdot g'=0$ for all broken fundamental weights $\lambda_i$, implying that  $g'=\sum_i m_i\alpha_i^*$  with only unbroken roots $\alpha_i$ and $m_i\geq0$. If $\{\alpha_i\}$ is a set of simple roots of $H\subset G$ then so is $\{\alpha'_i\}$ with $\alpha'_i=-\alpha_i$.  Since the Weyl group acts transitively on the bases of simple roots there exists an element in $W(H)$ that takes all unbroken roots $\alpha_i$ to $\alpha'_i=-\alpha_i$. With respect to the basis $(\alpha'_i)$ we have $g'= \sum_i m'_i \alpha'_i$ with $m'_i\leq 0$. This implies that $g'$ only satisfies the Murray condition if $g'=0$ showing that $g$ is indecomposable.\endproof 
We now wish to identify these topologically minimal charges. As a first step we  shall show that some of the coroots, that is roots of $G^*$ are contained in the set of topologically minimal charges. Note that there always exist coroots with topologically minimal charges, these correspond to the broken simple roots. If the residual symmetry group is non-abelian the set of topological minimal coroots is larger than the set of broken simple roots. In any case the whole set of topologically minimal coroots lies in the Murray cone.
\begin{prop}
\label{prop:mincoroot}
Any coroot $\alpha^*$ with $2\lambda_j\cdot \alpha^*=1$ for one of the broken fundamental weights and which is orthogonal to the other broken fundamental weights, satisfies the Murray condition.
\end{prop}
\proof We shall first show that $\alpha^*\cdot\mu \geq 0$. As argued in section \ref{subsect:murraycond} we take $\mu$ to lie in the closure of the fundamental Weyl chamber, i.e.~ $\mu= 2\sum_i\mu_i\lambda_i$ with $\mu_i\geq 0$. Thus $\alpha^*\cdot\mu= \mu_j\geq 0$.  If  $\alpha^*\cdot\mu=0$, $\alpha$ would be an unbroken root and as such orthogonal to all broken fundamental weights. This is clearly not the case since $2\lambda_j\cdot\alpha^*=1$. We conclude that $\alpha^*\cdot\mu>0$ and hence that $\alpha^*$ is a positive coroot.\\
It is now easy to show that $\alpha^*$ does indeed satisfy Murray's condition. Since the Weyl group is the symmetry group of the (co)root system we have for any $w\in W(H)\subset W(G)$, that $w(\alpha^*)$ is another coroot. Moreover $w(\alpha^*)$ is positive since the residual Weyl group leaves the Higgs VEV invariant: $w(\alpha^*)\cdot\mu= \alpha^*\cdot w^{-1}(\mu)= \alpha^*\cdot\mu$. We thus have that $w(\alpha^*)\cdot\mu>0$  for any $w\in W(H)$. Equaling some root of $G^*$ the positivity of $w(\alpha^*)$ implies  that it can be expanded in simple positive coroots with all coefficient greater than zero: $2\lambda_j\cdot w(\alpha^*)\geq 0$. We finally find that $2w(\lambda_i)\cdot\alpha^*\geq 0$ for all fundamental weights and  for all elements in the residual Weyl group.\endproof 
\begin{figure}[h!bt]
\centerline{
\includegraphics{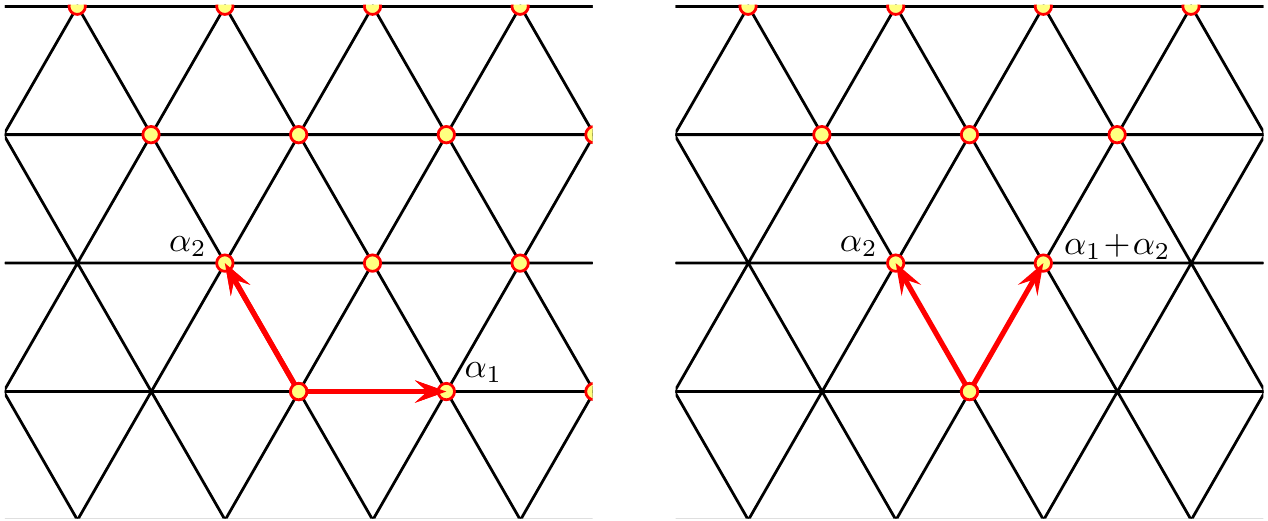}
}   
\caption{In the picture above the generators of the Murray cones of $SU(3)$ are depicted. If the gauge group is maximally broken (left) to $U(1)\times U(1)$, the generators correspond to the simple roots of SU(3). Both generating charges have distinct unit topological charges.  For minimal symmetry breaking (right) where the gauge group is $U(2)$, the Murray cone is further restricted by the Murray condition.  The generating magnetic charges do have distinct holomorphic charges related by the Weyl group. Their topological charges both equal 1.}
\label{fig:gensu3}

\end{figure}
It was easily shown that topologically minimal charges satisfying the Murray condition are indecomposable charges within the Murray cone. Furthermore we have seen that these topologically minimal charges contain the set of coroots with topologically minimal charges.  We will now prove that these coroots do not only constitute the complete set of minimal topological charges in the Murray cone, they actually form the full set of indecomposable charges. For $G= SU(3)$ these facts are easily verified in figure \ref{fig:gensu3} where the Murray cones and its generators are drawn for the two possible patterns of adjoint symmetry breaking.  Below we prove that the minimal topological charges generate the full Murray cone. Consequently the set of minimal topological charges must coincide with the complete set of indecomposable charges.   
\begin{prop}
The coroots with minimal topological charges generate the Murray cone
\end{prop}
\proof 
The outline of the proof is as follows. We slice up the Murray cone according to the topological charges in such  a way that each layer corresponds to a unique representation of the dual residual group. For unit topological charges we show that the weights correspond to the coroots with unit topological charges. Finally we show that the representations for higher topological charges pop up in the symmetric tensor products of  representations with unit topological charges.\\
Consider $G \to H$ where $H$ is locally of the form $U(1)^s\times K$.  We split the $r$ roots of the gauge group $G$ into $s$ broken roots $(\alpha_i)$ with $0< i \leq s$ and $r-s$ unbroken roots $(\alpha_j)$ with $ s <j \leq r$. The magnetic charges are thus expanded as $g= \sum_i m_i\alpha_i^* + \sum_j h_j \alpha_j^*$.\\
Without loss of generality we can assume $G$ to be simply connected just like in the proof of proposition \ref{prop:redhol}.  In that same proof we also defined an isomorphism $\lambda$ from the coroot lattice $\Lambda^*(G)$ to the weight lattice $\Lambda(H^*)$ of $H^*$. Since $H^*$ is locally of the form $U(1)^l\times K^*$ with $K$ semi-simple, $\lambda(g)$ can be expressed in terms of the $U(1)$ charges and a weight of $K^*$.  While the abelian charges are identified with the topological charges $m_i$, the Dynkin labels of the non-abelian charge are by $k_j= 2\alpha^*_j\cdot g/{\alpha_j^*}^2$. Being sums of  multiples of the entries of the Cartan matrix of $G^*$ these labels are indeed integers. Moreover for vanishing holomorphic charges only the off-diagonal entries contribute so that $k_j\leq 0$. Consequently for any $g\in \Lambda^*(G)$ we have:
\begin{equation}
\begin{split}
\lambda(g) &= \lambda\left(m_i\alpha_i^* + h_j\alpha_j^*\right) \\
           &=\lambda\left( m_i\alpha_i^* \right)+ \lambda\left( h_j \alpha_j^*\right)  \\
           &= h_-(m_i) + h_j\alpha_j^*.
\end{split}
\end{equation}
where $h_-(m_i)$ is a lowest weight that only depends on the topological charges.
We shall prove that for a fixed set of positive topological charges $\{m_i\}$ the magnetic charges in the Murray cone are in one-to-one relation with the weights  of the irreducible representation of $H^*$ labelled by $h_-(m_i)$. To show this we use two important facts. First a weight $h$ is in the representation defined by $h_-$ if and only if for the lowest weight $\tilde h $ in the Weyl orbit of $h$ one has $\tilde h = h_- + n_j\alpha_j^*$ where $n_j\geq 0$. Second, the map $\lambda$ commutes with the residual Weyl group.\\
First we shall show that for a magnetic charge $g$ in the Murray cone $\lambda(g)$ is a weight in the $h_-(m_i)$ representation. As a superficial consistency check we note that $\lambda(g)$ and $h_-(m_i)$ differ by an integer number of roots of $H^*$  given by the holomorphic charges. The lowest weight in the Weyl orbit of $\lambda(g)$ is given by the image of the reduced magnetic charge $\lambda(\tilde g)$, as explained in the proof of proposition \ref{prop:redhol}. It follows from the Murray condition (\ref{murraycond}) that $\tilde g$  is of the form $\tilde g =  m_i\alpha_i^* +  h''_j\alpha_j^*$ where $h''_j\geq 0$. Consequently $\lambda(\tilde g) = h_-(m_i) + n_j\alpha_j^*$ where $n_j \geq 0$.\\
To prove the converse we take a weight $h$ in the representation defined $h_-(m_i)$ with $m_i\geq 0$. We need to prove that $g$ with $\lambda(g)=h$ satisfies the Murray condition. This is done as follows. The triple  $(h_-(m_i),\tilde h, h)$ of weights in $\Lambda(H^*)$ can be mapped to a triple $(g_-(m_i), \tilde g , g)$ of elements in the coroot lattice  $\Lambda^*(G)$ by the inverse of $\lambda$. Next we show that $g_-(m_i), \tilde g$ and $g$ satisfy the Murray condition. We have $g_-(m_i)= m_i\alpha_i^*$ so that $\lambda(g_-)=\lambda_-(m_i)$ and $m_i\geq 0$. The broken simple coroots satisfy the Murray condition and hence $g_-(m_i)$ lies in the Murray cone. $\tilde g$ is given by $\tilde g = g_-(m_i)+ n_j\alpha_j^* $ so that $\lambda(\tilde g)= \lambda(g_-(m_i))+ n_j\alpha_j^* = \tilde h $.  Since $\tilde g$ maps to the anti-fundamental Weyl chamber of $H^*$ and has a positive expansion in simple coroots it satisfies the Murray conditions as follows from proposition \ref{prop:redhol}. Finally since $\lambda$ respects the residual Weyl group and $\tilde h$ is in the Weyl orbit of $h$ we find that $g$ is in the Weyl orbit of $\tilde g$. With $\tilde g$ satisfying the Murray condition it is easy to show that $g$ also obeys the condition.\\
The coroots of $G$ form the nonzero weights of the adjoint representation of $G^*$. Under symmetry breaking the adjoint representation maps to a reducible representation of $H^*$. We are particularly interested in the irreducible factors corresponding to unit topological charges. Coroots with unit topological charge, i.e.~$m_i = \delta_{ik}$,  equal a broken simple coroot $\alpha^*_k$ up to unbroken roots.  We have seen in proposition \ref{prop:mincoroot} that coroots with unit topological charge satisfy the Murray condition. Hence the previous discussion tells us that such coroots are mapped to the weight space of the representation labelled by $\lambda(\alpha^*_k)$. The weight $\lambda(\alpha^*_k)$ itself corresponds to $g=\alpha^*_k$. We now see that each weight in the $\lambda(\alpha^*_k)$-representation must not only correspond to a magnetic charge in the coroot lattice of $G$ but in fact to a coroot, otherwise the coroot system would not constitute a proper representation of $H^*$.\\
We can now finish the proof. Each element in the Murray cone is the weight in a representation labelled by $h_-(m_i)$. Such representations only depend on the topological charges. Moreover the lowest weights are additive with respect to the topological charges: $h_-(m_i)+h_-(m'_i)=h_-(m_i+m'_i)$. Consequently every such lowest weight is of the form $\sum_i m_i \lambda(\alpha^*_i)$. The representation labelled by $h_-(m_i) $ is obtained by the symmetric tensor product of representations labelled by $\lambda(\alpha_i)$. A weight in the product representation equals a sum of weights from the $\lambda(\alpha^*_i)$ representations. By identifying the weights with magnetic charges we find that all charges is the Murray cone equal a sum of coroots with unit topological charges.\endproof
\subsection{Generators of the magnetic weight lattice}
\label{section:genweight}
In this section we want to describe the generators of the magnetic charge lattice for singular monopoles in a theory with gauge group $H$. This charge lattice can be identified with the weight lattice $\Lambda(H^*)$  of the dual group $H^*$ as discussed in section \ref{sect:singcondition}. As for the Murray cone it is obvious that a minimal set of generating charges exists such that all charges are linear combinations of these generating charges with positive integer coefficients. The difference with the Murray cone however is that the generating set is not necessarily unique. We shall give some simple examples below to illustrate this, but we already note that the underlying reason for this is that the weight lattice of $H^*$ is closed under inversion.\\
\\
Using some textbook results on Lie group theory is easy to find a relatively small set of generators: let $V$ be a faithful representation of $H^*$ and $V^*$ its conjugate representation. 
Any irreducible representation of $H^*$ is contained in the tensor products of $V$ and $V^*$, see e.g section VIII of \cite{Coleman} for a proof. Since the weights of $V_1\otimes V_2$ are given by the sums of the weights of $V_1$ and $V_2$ we now find that any weight of an irreducible representation of $H^*$ is a linear combination of weights of $V$ and $V^*$ with positive coefficients. Since any weight in $\Lambda(H^*)$ is contained in an irreducible representation of $H^*$ we have found that the weights of $V$ and $V^*$ generate the magnetic weight lattice. Note that if this faithful representation $V$ is self-conjugate the weight lattice is obviously generated by the non-zero weights of $V$. This happens for example for $SO(n)$ and $Sp(2n)$ which have only self-conjugate representations. To find a small set of generators one should take the non-zero weights of a smallest faithful representation and its conjugate representation, i.e.~the fundamental representation and its conjugate representation.\\
The recipe above does not necessarily give a smallest set of generators since there still might be some double counting. We mention two examples. First $V^*$ might be contained in the tensor products of $V$. This happens for example for $SU(n)$: the representation $\mathbf{\bar n}$ is given by the $(n-1)$th anti-symmetric product of $\mathbf{n}$. Second some weights of $V$ may be decomposable within $V$. Consider for example $SU(n)/\mathbb Z_n$. The weight lattice of this group corresponds to the root lattice of $SU(n)$ and for $V$ one can take the adjoint representation whose weights are the roots of $SU(n)$. Note that all roots can be expressed as positive linear combinations of the simple roots and their inverses in the root lattice.\\
When $H^*$ is a product of groups the defining representation is reducible and falls apart into irreducible components. Each of these irreducible representations has trivial weights for all but one of the group factors. This agrees with the fact that in this case the weight lattice of $H^*$ is a product of weight lattices.\\ 
In table \ref{table:genwlat} we give the representation or representations whose nonzero weights constitute a minimal generating set of the magnetic weight lattice $\Lambda(H^*)$. The corresponding electric groups $H$ were mentioned in tables \ref{table:dualpairs} and \ref{table:dualpairs2}.
\begin{table}[!hbtp]
\begin{center}
\begin{tabular}{cc}
\hline
\hline
$H^*$ & $ \{V\}$ \\
\hline
$SU(n)$                           &  $\{\mathbf{n}\}$     \\
$Sp(2n)$                          &  $\{\mathbf{2n}\}$      \\
$SO(n)$                           &  $\{\mathbf{n}\}$ \\
$(U(1)\times SU(n))/\mathbb Z_n$  &  $\{\mathbf{n}_1$,  $\mathbf{\bar n}_{-1}\}$ \\
$ U(1)\times SO(2n+1)$            &  $\{(\mathbf{2n+1})_0$,  $\mathbf 1_1$, $\mathbf 1_{-1}\}$   \\
$(U(1)\times Sp(2n))/\mathbb Z_2$ &  $\{\mathbf{2n}_1$, $\mathbf{2n}_{-1}\}$  \\
$(U(1)\times SO(2n))/\mathbb Z_2$ &  $\{\mathbf{2n}_1$, $\mathbf{2n}_{-1}\}$  \\
\hline
\hline \\ 
\end{tabular}
\caption{Generators of the magnetic weights lattice $\Lambda(H^*)$ in terms of representations of the dual group $H^*$. The boldface numbers give the dimensionality of the irreducible representations of the corresponding simple Lie groups, their conjugate representations are distinguished by an extra bar. The subscripts denote $U(1)$-charges.}
\label{table:genwlat}
\end{center} 
\end{table}
\subsection{Generators of the fundamental Weyl chamber}
\label{subsect:genfundWeyl}
The charges of singular monopoles in a theory with gauge group $H$ take values in the weight lattice of the dual group $H^*$. This weight lattice has a natural subset: the weights in the fundamental Weyl chamber. If $H$ is semi-simple and has trivial center $H^*$ is semi-simple and is simply connected. In this particular case the generators of the fundamental Weyl chamber of $H^*$ are immediately identified as the fundamental weights. If $H^*$ is not simply connected or even not semi-simple the generating weights in the fundamental Weyl chamber are not that easily identified. The generating charges are however closely related to the generators of the representation ring, which are computed in chapter 23 of \cite{Fulton}. We shall explain this relation for the semi-simple, simply connected Lie groups. 
Finally we use the obtained intuition to compute the generators of the fundamental Weyl chamber for the dual groups in table \ref{table:dualpairs2} which occur in  minimal symmetry breaking of classical groups. In the next section we shall use similar methods to find the generators of the fundamental Murray cone.\\
\\
The representation ring $R(H^*)$ is the free abelian group on the isomorphism classes of irreducible representations of $H^*$. In this group one can formally add and subtract representations. The tensor product makes $R(H^*)$ into a ring. We shall for now assume $H^*$ to be a simple and simply connected Lie group of rank $r$ so that its weight lattice $\Lambda$ is generated by the $r$ fundamental weights $\{\lambda_i\}$.\\
$R(H^*)$ is isomorphic to a certain ring of Weyl-invariant polynomials. We will review the proof following \cite{Fulton}. We shall start by introducing $\mathbb Z[\Lambda]$, the integral ring on $\Lambda$. By this we mean that any element in $\mathbb Z[\Lambda]$ can be written as $\sum_\Lambda n_\lambda e_\lambda$ where $n_\lambda \in \mathbb Z$ and  $n_\lambda\neq 0$ for a finite set of weights. We thus see that $e_\lambda$ is the basis element in $\mathbb Z[\Lambda]$ corresponding to $\lambda$. The product in  $\mathbb Z[\Lambda]$ is defined by $e_\lambda e_{\lambda'} = e_{\lambda+\lambda'}$. We thus see that $\mathbb Z[\Lambda]$ is nothing but a group ring on the abelian group $\Lambda$. Note that the additive and multiplicative unit are given by $0$ and $e_0$ while the additive and multiplicative inverses of $e_\lambda$ are given by respectively $-e_\lambda$ and $e_{-\lambda}$.\\
\\
There is a homomorphism, denoted by $\text{Char}$, from the representation ring into $\mathbb Z[\Lambda]$ This map sends a representation $V$ to $\text{Char}(V)=\sum \text{dim} \left( V_\lambda\right) e_\lambda$, where $\text{dim} \left(V_\lambda\right)$ equals the multiplicity with which the weight $\lambda$ occurs in the representation $V$. It is easy to see that this map does indeed respect the ring structure.\\
The Weyl group $W$ of $H^*$ acts linearly on $\mathbb Z[\Lambda]$ and the action is defined by $w\in W: e_\lambda \mapsto e_{w(\lambda)}$. To show that the action of $W$ respects the multiplication in $\mathbb Z[\Lambda]$ one simply uses the fact that $W$ acts linearly on $\Lambda$.\\
$\mathbb Z[\Lambda]$ contains a subring $\mathbb Z[\Lambda]^W$ consisting of elements invariant under the Weyl group.  The claim is that $R(H^*)$ is isomorphic to $\mathbb Z[\Lambda]^W$. It is easy to show that the image of $\text{Char}$ is contained in $\mathbb Z[\Lambda]^W$. Below we shall also prove surjectivity by using the fact that there is a basis of $\mathbb Z[\Lambda]^W$ that is generated out of certain representation of $H^*$. In the end we are of course interested in these generators.\\
\\
To each dominant integral weight $\lambda\in \Lambda$ we associate an element $P_\lambda \in \mathbb Z[\Lambda]^W$ by choosing $P_\lambda = \sum n_{\lambda'}e_{\lambda'}$ with $n_{w(\lambda')} = n_{\lambda'}$ for all $w\in W $ and with $n_\lambda=1$.  For simplicity we take $P_\lambda$ so that $n_{\lambda'}=0$ if $\lambda-\lambda'$ is not a linear combination of roots. We now restrict the choice of $P_\lambda$ so that for any dominant integral weight $\lambda'>\lambda$, $n_{\lambda'}$ vanishes. Note that $\lambda$ is the highest weight of $P_\lambda$. One can now prove by induction that any set $\{P_\lambda\}$ satisfying the conditions above forms an additive basis for $\mathbb Z[\Lambda]^W$.\\
We shall now make a rather special choice for the basis $\{P_\lambda\}$. For the fundamental weights $\lambda_i$ we take $P_{\lambda_i}$ to be $P_i=\text{Char}(V_i)$ were $V_i$ is the irreducible representation of $H^*$ with highest weight $\lambda_i$. For any other dominant integral weight $\lambda=\sum m_i \lambda_i$ we take $P_\lambda = \text{Char}\left(\otimes_i V_i^{m_i}\right) = \Pi_i\, P_i^{m_i}$. Since $\{P_\lambda\}$ is a basis for $\mathbb Z[\Lambda]^W$ any element in this ring can thus be written as a polynomial in the variables $P_i$ with positive integer coefficients:
\begin{equation}
\mathbb Z[\Lambda]= \mathbb Z[P_1,\dots, P_r].
\end{equation}
\\
As promised we have proven that $R(H^*)$ is isomorphic to $\mathbb Z[\Lambda]^W$ for $H^*$ semi-simple and simply connected. In addition we have found that the generators of $\mathbb Z[\Lambda]^W$ correspond precisely to the generators of the fundamental Weyl chamber via the map $\lambda_i \mapsto P_i$. This is not very  surprising because it was input for the proof of the isomorphism. So the interesting question is if we can really retrieve the generators of the fundamental Weyl chamber from $R(H^*)$. This can indeed be done by identifying the generators of $\Lambda$ with the generators of $Z[\Lambda]$. We shall explain this below for $SU(n)$. Before we do so we want to make an important remark.\\
In the proof we used the fact that there is a basis ${P_\lambda}$ where each $P_\lambda$ can be identified with $\text{Char}(V_\lambda)$ and where $V_\lambda$ is some representation with highest weight $\lambda$. Such a choice of basis always exist since one can take $V_\lambda$ be the irreducible representation with highest weight $\lambda$. The fact that there is a generating set for the fundamental Weyl chamber is thus not crucial in the proof of the isomorphism between $R(H^*)$ and $\mathbb Z[\Lambda]^W$.\\
\\
We return to $\mathbb Z[\Lambda]$, where $\Lambda$ is the weight lattice of $SU(n)$. As discussed in the previous section the weight lattice of $SU(n)$ is generated by the weights of the $n$-dimensional fundamental representation. Let us denote these weights by $L_i$ and define
\begin{equation}
x_i =e_{L_i}\in \mathbb Z[\Lambda].
\end{equation}
Note that the vectors $L_i$ are not linearly independent since $\sum_i L_i =0$. We thus have
\begin{equation}
x_1x_2\cdots x_n= 1,
\end{equation} 
where $1=e_0$ is the multiplicative unit of $Z[\Lambda]$. We find that any element $e_\lambda$ can be written as monomial $\Pi_i\, x_i^{m_i}$ with positive coefficients $m_i$. Such monomials are unique up to factors $x_1\cdots x_n$. Since $\{e_\lambda : \lambda\in \Lambda\}$ forms a basis for $\mathbb Z[\Lambda]$ we find:
\begin{equation}
\mathbb Z[\Lambda]= \mathbb Z[x_1,\dots x_n]/(x_1\cdots x_n-1).
\end{equation}
The Weyl group of $SU(n)$ is the permutation group $\mathcal S_n$ and obviously permutes the indices of the $x_i$s. Consequently
\begin{equation}
R(SU(n)) = \mathbb Z[\Lambda]^{\mathcal S_n} = \mathbb Z[x_1,\dots x_n]^{\mathcal S^n}/(x_1\cdots x_n -1).
\end{equation}
To find the generators of $R(SU(n))$ we use the well known fact that any symmetric polynomial in $n$ variables can be expressed as a polynomial of  ${a_k: k=1,\dots ,n}$ where $a_k$ is the $k$th elementary symmetric function of $x_i$ given by:
\begin{equation}
a_k = \sum_{i_1< \cdots < i_k} x_{i_1}\cdots x_{i_k}.
\end{equation}
Note that $a_n=x_1\cdots x_n$ is identified with $1$ in $R(SU(n))$. We have thus established the isomorphism:
\begin{equation}
R(SU(n)) = \mathbb Z[a_1,\dots, a_{n-1}].
\end{equation}
Our conclusion is that the first $n-1$ elementary symmetric functions form a minimal set generating the representation ring of $SU(n)$. It should not be very surprising that for $i<n$ $a_i= P_i = \text{Char}(V_i)$ where $V_i$ is the irreducible representation with highest weight $\lambda_i$. It is nice to note that $V_i = \wedge^i V$ where $V$ is the fundamental representation of $SU(n)$ and that $\wedge^n V =1$ the trivial representation.\\
\\
For $SO(2n+1)$, $Sp(2n)$, and $SO(2n)$ the fundamental representation has $2n$ nonzero weights ${\pm L_i: i=1,\dots n}$. By identifying $x_i^{\pm 1} = e_{\pm L_i}$ one finds that the group ring on the weight lattice is isomorphic to $\mathbb Z[x_1,x_1^{-1},\dots,x_n,x_n^{-1}]$. As shown in \cite{Fulton} the representation rings are given by polynomial rings of the form:
\begin{align}
R(SO(2n+1))&= \mathbb Z[b_1,\cdots, b_n]  \\
R(Sp(2n))&= \mathbb Z[c_1,\cdots, c_n] \\
R(SO(2n))&= \mathbb Z[d_1,\cdots , d_{n-1}, d_n^+, d_n^-].
\end{align}
The polynomials $b_k$, $c_k$ and $d_k$ can all be chosen to equal the elementary symmetric functions in the $2n$ variables $\{x_i^{\pm}\}$. The polynomials $d_n^\pm$ can be expressed as $(d^\pm)^2$. $d^+$ and $d^-$ correspond to the two spinor representations of $SO(2n)$ :
\begin{equation}
\label{eqn:spinchar}
d^\pm = \text{Char}(S_{\pm}) = \sum_{s_1\cdots s_n= \pm 1} \sqrt{x_1^{s_1}\cdots x_n^{s_n}}.
\end{equation} 
It is easy to check that $d_n^\pm$ are indeed polynomials.\\
To explain why $R(SO(2n))$ has an extra generator compared to the other groups we note that its Weyl group is given by $\mathcal S_n \ltimes \mathbb Z_2^{n-1}$ whereas the Weyl groups of $SO(2n+1)$ and $Sp(2n)$ are given by $\mathcal S_n \ltimes \mathbb Z_2^n$. This means that the Weyl groups act on the non-zero weights of the fundamental representations by permuting the indices and changing the signs of the weights, but for $SO(2n)$ only an even number of sign changes is allowed. Consequently the generators of $R(SO(2n))$ do not have to be invariant under for example of $x_1 \mapsto x_1^-$ and hence the generator $d_n$ can be decomposed into $d_n^{+}$ and $d_n^{-}$.\\
for completeness we mention that the highest weights of $b_k$, $c_k$ and $d_k$ are given by the highest weights of the anti-symmetric tensor products  $\wedge^k V$ of the corresponding fundamental representation $V$. The highest weights of $d_n^{\pm}$ are given by twice the highest weight of the spinor representations $S^\pm$.\\
\\
We finally want to identify the generators of the fundamental Weyl chamber for some groups that arise in minimal symmetry breaking of classical groups. 
As discussed in section \ref{section:genweight} the weight lattice $\Lambda$ of $U(n)$ is generated by the weights of its $n$-dimensional representation $\mathbf n_1$ and those of its conjugate representation $\mathbf{\bar n}_{-1}$. Let us denote the weights of $\mathbf n_1$ by $\{L_i\}$ and define $x_i=e_{L_i}\in \mathbb Z[\Lambda]$. The weights of $\mathbf{\bar n}_{-1}$ are given by $\{-L_i\}$. We thus immediately find the following isomorphism for the group ring on the weight lattice of $U(n)$:
\begin{equation}
\mathbb Z[\Lambda] = \mathbb Z[x_1,x_{-1},\dots,x_n,x_n^{-1}].
\end{equation} 
To find the generators of the representation ring $R(U(n)) = \mathbb Z[\Lambda]^W$ we note that the Weyl group $W = \mathcal S_n$ of $U(n)$ permutes the indices of the generators of $\mathbb Z[\Lambda]$ but does not change any of the signs as happened for the classical groups discussed right above. This implies that $R(U(n))$ is generated by $\{a_k: k=1,\dots, n\}$ the elementary symmetric polynomials in ${x_i}$ and $\{\bar a_k: k=1,\dots, n\}$ the elementary symmetric polynomials in the variables $\{x_i^{-1}\}$. Note that $a_n = x_1\cdots x_n$ is invertible in the representation ring and its inverse is given by $\bar a_n = (x_1\cdots x_n)^{-1}$.\\
The generators we have found for $R(U(n))$ are not completely independent since:
\begin{equation}
a_ka_n^{-1}= \!\!\!\!\sum_{i_{j-1}< i_j < i_{j+1}}\!\!\!\!\!\!\! x_{i_1}\cdots x_{i_k}(x_1\cdots x_n)^{-1} = \!\!\!\!\sum_{i_{j-1}< i_j < i_{j+1}}\!\!\!\!\!\!\! (x_{i_1}\cdots x_{i_{n-k}})^{-1} = \bar a_{n-k}.
\end{equation} 
The representation ring of $U(n)$ can thus be identified with the polynomial ring:
\begin{equation}
\label{eqn:repringUn}
R(U(n))=\mathbb Z[a_1,\dots, a_n, a_n^{-1}].
\end{equation}
The generating polynomials $a_k$ and $a_n^{-1}$ are indecomposable in the representation ring, their highest weights thus form a minimal set generating the fundamental Weyl chamber of $U(n)$. We finally mention that  $a_k = \text{Char}(\wedge^k V)$, where $V$ is the fundamental representation of $U(n)$. Moreover $\wedge^n V$ is the one dimensional representation that acts by multiplication with $\text{det}(g)$ where $g\in (U(n))$ This representation is invertible and $a_n^{-1} =\text{Char}\left((\wedge^n V)^{-1})\right)$.\\
\\
Since $U(1)\times SO(2n+1)$ is a product of groups its representation ring is simply $R(U(1))\times R(SO(2n+1)$. The representation ring of $U(1)$ can be identified with the polynomial ring $\mathbb Z[x_0,x_0^{-1}]$ where $x_0^{\pm 1}= \text{Char}(V^{\pm 1})$  and $V$ the fundamental representation of $U(1)$. There is however an alternative description of the representation ring which will prove to be valuable in the next section. Let $\{L_0,L_1,L'_1,\dots, L_n,L'_n\}$ be the weights of the fundamental representation of $U(1)\times SO(2n+1)$, i.e.~the representation with unit $U(1)$ charge. Define $\{x_0,x_1,{x'}_1,\dots x_n,x_n'\}$ to be the images of these weights in the group ring $\mathbb Z[\Lambda]$ of the weight lattice. It is not too hard to show that $\mathbb Z[\Lambda]$ is isomorphic to  $\mathbb Z[x_0, x_0^{-1},x_1,{x'}_1,\dots x_n,x_n']/I$ where $I$ is the ideal generated by the relations  $x_i{x'}_i= x_0^2$. Moreover one can prove that 
\begin{equation}
R(U(1)\times SO(2n+1)) = \mathbb Z[x_0,x_0^{-1},b_1,\dots, b_n],
\end{equation}  
where $b_k=\text{Char}(\wedge^k V)$ is the $k$th elementary symmetric polynomial in the $2n+1$ variables $x_0,x_1,{x'}_1,\dots, x_n,{x'}_n$. The highest weights of these generating polynomials correspond to the minimal set of generating charges in the fundamental Weyl chamber.\\
\\
By mapping the weights of the fundamental representation and its conjugate representations to $\mathbb Z[\Lambda]$ on finds that group rings on the weight lattices of $(U(1)\times Sp(2n))/\mathbb Z_2$ and $(U(1) \times SO(2n))/\mathbb Z_2$ can be identified with the polynomial ring
\begin{equation}
\mathbb Z[x_1,x_1^{-1},{x'}_1, {x'}_1^{-1},\dots, x_n,x_n^{-1},x_n', {x'}_{n-1}]/I,
\end{equation} 
where $I$ is the ideal generated by the relations $x_i{x'}_i= x_j{x'}_j$. Note that these relations imply that $x_1x'_1$ is invariant under the Weyl group that permutes the indices and swaps primed variables with their unprimed counterparts. One can now show that the representation rings $R\left((U(1)\times Sp(2n))/\mathbb Z_2\right)$ and $R\left((U(1)\times SO(2n))/\mathbb Z_2\right)$  can be identified as quotient rings of respectively:
\begin{equation}
\mathbb Z[x_1{x'}_1, (x_1{x'}_1)^{-1}, c_1,\dots, c_n, \bar c_1,\dots, \bar c_n]
\end{equation}
and
\begin{equation}
\mathbb Z[x_1{x'}_1, (x_1{x'}_1)^{-1}, d_1,\dots, d_{n-1}, d_n^+, d_n^-, \bar d_1,\dots, \bar d_{n-1}, \bar d_n^+, d_n^-],
\end{equation}
where $c_k$ and $d_k$ are the elementary symmetric polynomials in the $2n$ variables $x_1,\dots, x_n$ and $x'_1,\dots, x'_n$. The functions $\bar c_k$ and $\bar d_k$ are similar elementary symmetric polynomials expressed in terms of the inverted variables. Explicit expressions for  $d_n^\pm= (d^\pm)^2$ can be found from formula (\ref{eqn:spinchar}) where the inverted variables should be replaced by the primed variables. Finally $\bar d_n^\pm$ is found by substitution of the inverted variables in $d_n^\pm$. The generating set of polynomials we have found is not the minimal set. This follows from the fact that $x_j^{-1}= (x_jx'_j)^{-1}x'_j= (x_1x'_1)^{-1}x'_j$. Consequently one finds:
\begin{equation}
R\left((U(1)\times Sp(2n))/\mathbb Z_2\right) =\mathbb Z[x_1{x'}_1, (x_1{x'}_1)^{-1}, c_1,\dots, c_n]
\end{equation}
\begin{equation}
R\left((U(1)\times SO(2))/\mathbb Z_2\right) =  \mathbb Z[x_1{x'}_1, (x_1{x'}_1)^{-1}, d_1,\dots, d_{n-1}, d_n^+, d_n^-].
\end{equation}
The highest weights of these generating polynomials  are the generators of the fundamental Weyl chamber of the two groups.
\subsection{Generators of the fundamental Murray cone}
\label{subsect:genfundcone}
The fundamental Murray cone, just like the Murray cone, contains a unique set of indecomposable charges. The uniqueness of this set is a consequence of the fact that the fundamental Murray cone does not allow for invertible elements. The main difference with the Murray cone however is that the generators for the fundamental Murray cone are not easily computed. After a general discussion we shall therefore only  determine the generators for a couple of cases that correspond to minimal symmetry breaking of classical groups. The approach we use is closely related to the computation of the generators of the fundamental Weyl chamber as discussed in the previous section and can in principle be applied to any gauge group and for arbitrary symmetry breaking.\\
\\
Note that this whole exercise only makes sense if the fundamental Murray cone is closed under addition. At the beginning of section \ref{subsect:genmurray} we argued that the Murray cone is closed under this operation by evaluating the defining equations. For the fundamental Murray cone similar considerations apply. For $g$ to be in the fundamental Weyl chamber of the Murray cone we have the extra condition $g\cdot\alpha_i \geq 0$ for all unbroken roots $\alpha_i$. It is now easily seen that if both $g$ and $g'$ satisfy this condition then $g+g'$ will satisfy it too, as will any linear combination of these charges with positive integer coefficients. This proves that the fundamental Murray is closed under addition of charges.\\
\\
Instead of computing the generators of the fundamental Murray cone directly by evaluating the Murray condition we shall determine the indecomposable generators of a certain representation ring. We shall start by describing this ring. Let $G$ be a compact, semi-simple group broken to $H$ via a adjoint Higgs field. Without loss of generality we can assume $G$ to be simply connected since this does not change the set of magnetic charges. Under this condition the magnetic weight lattice $\Lambda := \Lambda(H^*)$ is isomorphic to the root lattice of $G^*$.  The ring we want to consider is the free abelian group on the irreducible representations of $H^*$ with weights in the Murray cone. These irreducible representations of $H^*$ are labelled by dominant integral weights in $\Lambda_+\subset \Lambda$ and can be identified with the fundamental Murray cone as a set.  Note that since the Murray cone is closed under addition this set of representations is closed under the tensor product.  As we proof in the appendix there exists an algebraic object, but not a group, having  a complete set of irreducible representations labelled by the magnetic charges in the Murray cone. Let us denote this object by $H^*_+$. The representation ring we are discussing here is thus precisely the representation ring $R(H^*_+)$.\\  
\\
Just as in the previous section we now introduce a second ring $\mathbb Z[\Lambda_+]$  that turns out to be quite useful. $\mathbb Z[\Lambda_+]$ has a basis $\{e_\lambda : \lambda\in \Lambda_+\}$. Since $\Lambda_+$ is closed under addition $\mathbb Z[\Lambda_+]$ is indeed closed under multiplication. The multiplicative identity is given by $1=e_0$. The basis elements $e_\lambda$ of $\mathbb Z[\Lambda_+]$ are not invertible under multiplication since $e_{-\lambda}$ is not contained in  $\mathbb Z[\Lambda_+]$. Finally we introduce the ring $\mathbb Z[\Lambda_+]^W$ consisting of the Weyl invariant elements in $\mathbb Z[\Lambda_+]$. Note that $\mathbb Z[\Lambda_+]\subset \mathbb Z[\Lambda]$ and $\mathbb Z[\Lambda_+]^W \subset \mathbb Z[\Lambda]^W$. By using arguments almost identical to arguments mentioned in the previous section one can show that $R(H^*_+)$ is isomorphic to $\mathbb Z[\Lambda_+]^W$. This last ring can be identified with a polynomial ring. The highest weights of the indecomposable polynomials can be identified with the generators of the fundamental Murray cone. \\
\\
We shall identify the generators of the fundamental Murray cone for the classical simply-connected groups $SU(n+1)$, $Sp(2n+2)$, $Spin(2n+3)$,  and $Spin(2n+2)$ and for minimal symmetry breaking.  The relevant residual electric groups and their magnetic dual groups are listed in table \ref{table:dualpairs2}. One can show that the Murray cone in these cases is generated by the weights of the fundamental representation of $H^*$ which are respectively $n$, $2n+1$, $2n$ and $2n$ dimensional.\\
\\
Let us denote the weights of the fundamental representation of $U(n)$ by $L_i$ where $i=1,\dots,n$. We define $x_i=e_{L_i}$. Since the weights $L_i$ freely generate the Murray cone we immediately find
\begin{equation}
\mathbb Z[\Lambda_+] = \mathbb Z[x_1,\dots, x_n].
\end{equation}  
The Weyl group of $U(n)$ permutes the indices of the generators. Copying our results of the previous section we thus find the following isomorphism:
\begin{equation}
\mathbb Z[\Lambda_+]^W = \mathbb Z[a_1,\dots, a_n].
\end{equation}
where $a_k$ are the elementary symmetric polynomials in the variables $x_i$. The highest weights of these indecomposable polynomials are the generators of the fundamental Murray cone for $SU(n+1)$ broken down to $U(n)$. Note that $\mathbb Z[\Lambda_+]^W$ is obtained from  $\mathbb Z[\Lambda]^W$ as given in formula (\ref{eqn:repringUn}) by removing the generator $a_n^{-1}$.\\
\\
Let $\{L_0,L_1,L'_1,\dots, L_n,L'_n\}$ be the weights of the fundamental representation of $U(1)\times SO(2n+1)$. Define $\{x_0,x_1,{x'}_1,\dots x_n,x_n'\}$ to be the images of these weights in the ring $\mathbb Z[\Lambda_+]$.  $\mathbb Z[\Lambda_+]$ is isomorphic to 
\begin{equation}
\mathbb Z[x_0,x_1,{x'}_1,\dots x_n,x_n']/I,
\end{equation}
 where $I$ is the ideal generated by the relations  $x_i{x'}_i= x_0^2$. Moreover one can now prove that 
\begin{equation}
\mathbb Z[\Lambda_+]^W  = \mathbb Z[x_0,b_1,\dots, b_n],
\end{equation}  
where $b_k=\text{Char}(\wedge^k V)$ is the $k$th elementary symmetric polynomial in the $2n+1$ variables $x_0,x_1,{x'}_1,\dots, x_n,{x'}_n$. The highest weights of these generating polynomials correspond to the minimal set of generating charges in the fundamental Murray cone.\\
\\
By mapping the weights of the fundamental representation to $\mathbb Z[\Lambda_+]$ for $G$ equals $Sp(2n+2)$ or $SO(2n+2)$ the ring  $\mathbb Z[\Lambda_+]$ can be identified with the polynomial ring
\begin{equation}
\mathbb Z[x_1,{x'}_1,\dots, x_n,x_n']/I, 
\end{equation} 
where $I$ is the ideal generated by the relations $x_i{x'}_i= x_j{x'}_j$. One can now show that the representation rings can be identified as respectively:
\begin{equation}
\mathbb Z[x_1{x'}_1, c_1,\dots, c_n]
\end{equation}
and
\begin{eqnarray}
\mathbb Z[x_1{x'}_1, d_1,\dots, d_{n-1}, d_n^+, d_n^-],
\end{eqnarray}
where $c_k$ and $d_k$ are both elementary symmetric polynomials in the variables $x_1,\dots, x_n$ and $x'_1,\dots, x'_n$.  Explicit expressions for  $d_n^\pm= (d^\pm)^2$ are the same as the corresponding generating polynomials for $Z[\Lambda]^W$. The generators of the fundamental Murray cone can be found by computing the highest weights of the polynomials. 
\section{Moduli spaces for smooth BPS monopoles}
\label{sect:modspace}
For both singular and smooth monopoles we have identified the set of magnetic charges. This set always contains a subset closed under addition that arises by modding out Weyl transformations. On top of this we have seen that these sets are generated by a finite set of magnetic charges. This suggest that these generating charges correspond to a distinguished collection of basic monopoles and that all remaining magnetic charges give rise to multi-monopole solutions. By studying the dimensions of moduli spaces of solutions we can try to confirm this picture. In this section we shall only be concerned with smooth BPS monopoles. For such monopoles the magnetic charges satisfy the Murray condition.
\subsection{Framed moduli spaces}
\label{subsection:fulframed}
The moduli spaces we shall discuss in this section are so-called framed moduli spaces. Such spaces are commonly used in the mathematically oriented literature on monopoles, see, for example, the book  \cite{Atiyah:1988jp}. We shall discuss these spaces presently. In the next sections we review the counting of dimensions. \\
\\
The moduli spaces we are considering correspond to a set of BPS solutions modded out by gauge transformations. The set of BPS solutions is restricted by the boundary condition we use, as discussed in section \ref{subsect:quantBPS}.
Beside the finite energy condition one can use additional framing conditions, hence the terminology framed moduli spaces. 

Recall from our discussion following \eqref{eqn:boundcond} that
the value $\phi(\hat r_0)$ of the asymptotic Higgs field at an arbitrarily
chosen point $\hat r_0$ on the two-sphere at infinity 
 determines the residual gauge group. It is therefore 
natural to restrict the configuration space to BPS solutions with 
$\phi(\hat r_0)=\Phi_0$ for a fixed value of $\Phi_0$. 
The resulting space has multiple connected components labelled by the topological charge of the BPS solutions. This topological charge is given by the topological components $m_i$ of $G_0=G(\hat r_0)$ as explained in section \ref{sect:chargelat}. We shall thus consider the finite energy configurations satisfying the framing condition
\begin{equation}
\label{eqn:framecond}
\Phi(t\hat r_0)= \Phi_0 - \frac{G_0}{4\pi t}+\mathcal O \left(t^{-(1+\delta)}\right) \qquad t \gg 1,
\end{equation}  
where $\hat r_0$, $\Phi_0$ and the topological components $m_i$ of $G_0$ are completely fixed. 
The framed moduli space $\mathcal M(\hat r_0,\Phi_0, m_i)$ is now obtained from the configuration space by modding out certain gauge transformations that respect the framing condition. The full group of gauge transformations $\mathcal G: \mathbb R^3 \to G$ that respect this condition satisfy $\mathcal G(t\hat r_0)=h$ as $t\to \infty$  where $h\in H$. However, for the moduli space to be a smooth manifold one can only mod out a group of gauge transformations that acts freely on the configuration space. For example the configuration $\Phi=\Phi_0$ and $B=0$ is left invariant by all constant gauge transformations given by $h\in H$. The framed moduli space is thus appropriately defined as the space of BPS solutions satisfying the boundary conditions (\ref{eqn:boundcond}) and (\ref{eqn:framecond}), modded out by the gauge transformations that become trivial at the chosen base point $\hat r_0$ on the sphere at infinity.\\
\\
The moduli space $\mathcal M(\hat r_0, \Phi_0, m_i)$ has several interesting subspaces which will play an important role in what is to come. These subspaces are related to the fact that there is a map $f$ from the moduli space to the Lie algebra of $G$. This map is defined by assigning $G_0$ to each configuration. As explained in section \ref{subsect:quantBPS} and \ref{subsect:murraycond}, up to a residual gauge transformation $G_0$ is given by  $G_0= \frac{4\pi}{e}g\cdot H$  with $g$ an element in the fundamental Murray cone. The topological components of $g$ are of course fixed while the holomorphic charges are restricted by the topological charges. The image of $f$ in the Lie algebra of $G$ is thus a disjoint union of $H$ orbits
\begin{equation}
C(g_1)\cup \cdots \cup C(g_n),
\end{equation}
where $g_i$ is the intersection of each orbit with the fundamental Murray cone. The map $f$ defines a stratification of  $\mathcal M(\hat r_0, \Phi_0, m_i)$. Each stratum $\mathcal M_{g_i}$ is mapped to a corresponding orbit $C(g_i)$ in the Lie algebra.\\
The remarkable thing about the stratification is that for a fixed topological charge the strata are disjoint but connected even though the images of the strata are disconnected sets in the Lie algebra of $G$. This follows from the fact that all BPS configurations in  $\mathcal M(\hat r_0, \Phi_0, m_i)$ are topologically equivalent and can be smoothly  deformed into each other. Under such smooth deformations the holomorphic charges can thus jump.\\
If the residual gauge group is abelian the stratification is trivial. Since the topological charges completely fix $g$ there is only a single stratum $\mathcal M_g=\mathcal M(\hat r_0, \Phi_0, m_i)$.\\ 
\\
There is another interesting moduli space we want to introduce.  This so-called fully framed moduli space $\mathcal M(\hat r_0, \Phi_0, G_0) \subset \mathcal M(\hat r_0, \Phi_0, m_i)$ arises by imposing even stronger framing conditions. The points in the fully framed moduli space  $\mathcal M(\hat r_0, \Phi_0, G_0)$ correspond to BPS configurations obeying the usual boundary conditions (\ref{eqn:boundcond}) and (\ref{eqn:framecond}) but instead of only fixing  $\Phi_0$ we also choose a completely fixed magnetic charge $G_0$. Again the gauge transformations that become trivial at the chosen base point are modded out.\\
The fully framed moduli spaces have a special property in relation to the strata. Monopoles with magnetic charges $G'_0$ and $G_0$ related by $h$ in residual gauge group $H\subset G$ lie in the same stratum of the framed moduli space. Moreover, the action of $h\in H\subset G$ on the magnetic charges can be lifted to a gauge transformation $\mathcal G: S^2 \to G$ \cite{Murray:2003mm}. Since $\pi_2(G)=0$ this gauge transformation can in turn be extended to a gauge transformation in $\mathbb R^3$ acting on the complete BPS solution. In other words the action of $h\in H$ on the Lie algebra  can be lifted to an action on the framed moduli space such that each point in $\mathcal M(\hat r_0, \Phi_0, G_0)$ is mapped to a point in $\mathcal M(\hat r_0, \Phi_0, G_0)$. We thus see that all fully framed moduli spaces in a single stratum are isomorphic. In addition we also have that a stratum is nothing but a space of fully framed moduli spaces. Finally we conclude that locally we must have that $M_i = C(g_i)\times M(\hat r_0, \Phi_0, G_0)$ where $G_0$ is defined by $g_i$.\\
If the residual gauge group is abelian the action of $H$ on the magnetic charges is trivial. In this particular case the fully framed moduli space equals the single stratum and we have $\mathcal M(\hat r_0, \Phi_0, G_0) = \mathcal M(\hat r_0, \Phi_0,m_i)$. 
\subsection{Parameter counting for abelian monopoles}
\label{sect:abelian}
\begin{figure}[!hbtp]
\centerline{
\includegraphics{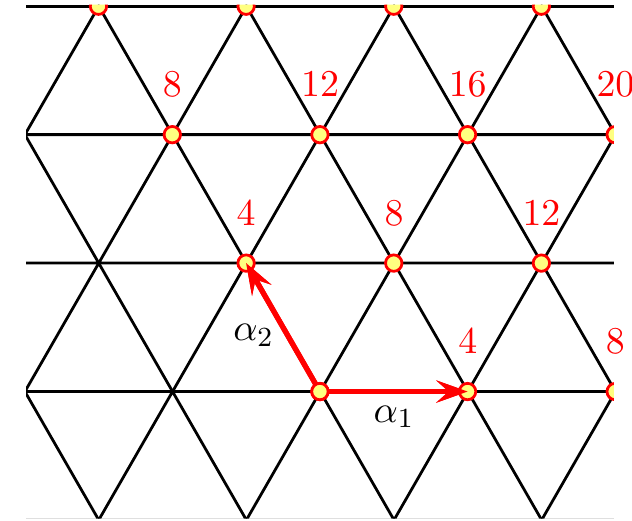}
}   
\caption{The Murray cone for $SU(3)$ broken to  $U(1)\!\times \!U(1)$. The generators of the cone are precisely the simple (co)roots $\alpha_1$ and $\alpha_2$ of SU(3). Both these charges correspond to unit topological charge in $\pi_1(U(1)^2)=\mathbb Z\!\times \!\mathbb Z$. All charges can be decomposed into the generating charges. The dimensions of the moduli spaces are proportional to the number of components. These dimensions are obviously additive.}
\label{fig:dimabel}
\end{figure}

The dimensions of the framed moduli spaces for maximal symmetry breaking have been computed by Erick Weinberg \cite{Weinberg:1979zt}. From his index computation Weinberg concluded that there must be certain fundamental monopoles and that the remaining  monopoles should be interpreted as multi-monopole solutions. The magnetic charges of these fundamental monopoles are precisely the generators of the Murray cone. Note that since there is no distinction between the Murray cone and the fundamental Murray cone in the abelian case we may also call these fundamental monopoles basic. Our conclusion is thus that the moduli space dimensions are consistent with the structure of the (fundamental) Murray cone. This result also holds in the non-abelian case albeit in a much less obvious way. To get some feeling for this general case we shall first briefly review Weinberg's results.\\
\\
As before we consider a Yang-Mills theory with a gauge group $G$. The adjoint Higgs VEV $\mu$ is taken such that the gauge group is broken to its maximal torus $U(1)^r$, $r$ is the rank of the group.  In this abelian case  the structure of the framed moduli as well as the structure of the Murray cone is relatively simple. Since there is no residual non-abelian symmetry there are no holomorphic charges. Consequently the magnetic charge is fully determined by the topological charges and the action of the residual gauge group on the magnetic charges is trivial. The fully framed moduli spaces thus coincide with the framed moduli spaces while the fundamental Weyl chamber of the Murray cone is identical to the complete cone. From the Murray-Singer analysis it follows that the stable magnetic charges are of the form:
\begin{equation}
\label{eqn:gexp}
g= \sum_{i=1}^r m_i\alpha^*_i~, ~~m_i\in \mathbb N.
\end{equation}   
The $r$ simple coroots $\alpha_i^*$ obviously generate the Murray cone and the positive expansion coefficients $m_i$ can be identified with the topological charges as explained in section \ref{subsect:quantBPS}.\\
According to the index calculations of Weinberg the dimensions of the moduli spaces are proportional to the topological charge:
\begin{equation}
\label{eqn:dimabel}
\text{dim }\mathcal M_g= \sum_{i=1}^r 4m_i.
\end{equation}
As an illustration the Murray cone is depicted for $SU(3)\to U(1)^2$  in figure \ref{fig:dimabel}. For each charge the dimension of the moduli space is given. In general there are $r$ indecomposable charges, one for each $U(1)$-factor. These basic monopoles all have unit topological charge. Thus we see that the dimension of the moduli space is proportional to the  number $N= \sum m_i$ of indecomposable charges constituting the total charge. As Weinberg concluded this is precisely what one would expect for $N$ non-interacting monopoles, and hence is seems consistent to view the higher topological charge solutions as multi-monopole solutions. \\
\\
Before we continue with general symmetry breaking let us  pause for moment to discuss the nature  of the moduli space dimensions. These dimensions correspond to certain parameters of the BPS solutions. For the basic monopoles with charge $\alpha_i^*$ the obvious candidates for three of these are their spatial coordinates, i.e.~the position of the monopole. The fourth is related to electric action by $H_{\alpha_i}$ which keeps the magnetic charge fixed but nevertheless acts non-trivially on the monopole solutions. 
This can be seen by considering exact solutions for the basic monopoles obtained by embedding $SU(2)$ monopoles \cite{Bais:1978fw,Weinberg:1979zt}. \\
If the multi-monopole picture is correct the nature of the moduli space dimensions for higher topological charge is easy to guess. $3N$ correspond to the positions of the $N$ constituents, while the remaining $N$ dimensions arise from the action of the gauge group on the constituents. It has been shown by Taubes \cite{Taubes:1981gw} that if $\sum m_i = N$  there exists an exact BPS solutions corresponding to $N$ monopoles with unit topological charges. A similar result was obtained by Manton for two 't Hooft-Polyakov monopoles \cite{Manton:1977er}. The positions of the individual monopoles can be chosen arbitrarily as long as the monopoles are well separated. This immediately confirms the given interpretation of the $3N$ parameters. Further evidence for this interpretation of the moduli space parameters can be found by studying the geodesic motion on the moduli space.  For $N$ widely separated monopoles the geodesic motion on the asymptotic moduli space corresponds to the motion of $N$ dyons, considered as point-particles in $\mathbb R^3$, interacting via Coulomb-like forces. The conserved electric $U(1)$-charges appear in the geodesic approximation on the asymptotic moduli space because the metric has $U(1)$-symmetries. The correspondence between the classical theory on the asymptotic moduli space and the effective theory of classical dyons in space has up till now only been demonstrated for an arbitrary topological charge in a $SU(n)$-theory broken to $U(1)^n$ \cite{Manton:1985hs,Gibbons:1995yw,Bielawski:1998hj,Bielawski:1998hk,Bak:1997pc} and for topological charge 2 in an arbitrary theory with maximal symmetry breaking \cite{Lee:1996kz}.
\subsection{Parameter counting for non-abelian monopoles}
\label{sect:stratadims}
\begin{figure}[!ht]
\centerline{
\includegraphics{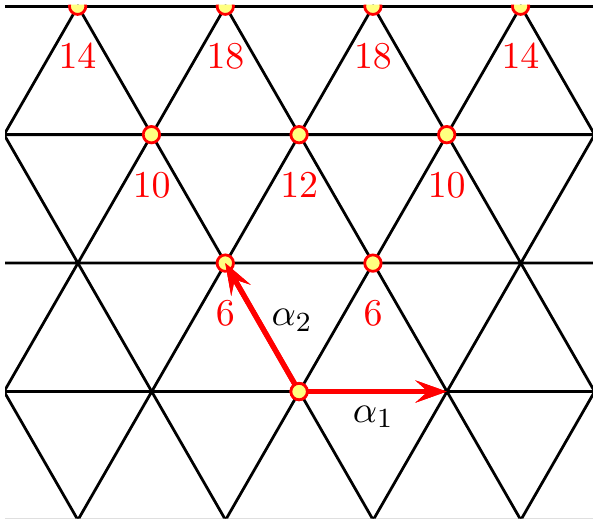}
}   
\caption{Dimensions for the strata of the framed  moduli spaces for $SU(3)$ broken to $U(2)$.}
\label{fig:stratadimsu2}
\end{figure}

Just as in the abelian case the dimensions of the framed moduli spaces for non-abelian mono\-poles are proportional to the topological charges. Hence the dimensions of the moduli spaces respect the addition of charges in the Murray cone. In that sense one could once more interpret monopoles with higher topological charges as multi-monopole solutions built out of monopoles with unit topological charges. This analysis would however ignore the fact that both the framed moduli space and the Murray cone have extra structure. The framed moduli space has a stratification while the magnetic charges have topological and holomorphic components. The holomorphic charges and thereby the strata are physically very important because they are directly related to the electric symmetry that can be realized in the monopole background as we shall discuss later in the section. Therefore one should wonder if these structures are compatible and if so how they will affect the multi-monopole interpretation.\\ 
\\
The dimensions for the framed moduli spaces of monopoles have been computed by Murray and Singer for any possible residual gauge symmetry, either abelian or non-abelian \cite{Murray:2003mm}. Their computation does not rely on index methods but instead it is based  on the fact that framed moduli spaces can be identified with certain sets of rational maps. Such a bijection was first proved by Donaldson for $G=SU(2)$ \cite{Donaldson:1985id} and later generalized by Hurtubise and Murray for maximal symmetry breaking \cite{Hurtubise:1989qy,Hurtubise:1989wh,Hurtubise:1990zf}. Finally the correspondence between framed moduli spaces and rational maps was proved for general gauge groups and general symmetry breaking by Jarvis \cite{Jarvis1,Jarvis2}. 
Murray and Singer have computed the dimensions of these spaces of rational maps. For further details we refer to the original paper. The $SU(n)$ case can also be found in \cite{Murray:1989zk}.\\
\\
One of the results of the calculations in \cite{Murray:2003mm} is that the dimension of the framed moduli space $\mathcal M(\hat r_0,\Phi_0,m_i)$ is given by:
\begin{equation}
\label{eqn:dimnonabel}
\text{dim }\mathcal M (\hat r_0, \Phi_0, m_i)= 4\sum_{i=1}^s\left(1- 2\rho\cdot \alpha^*_i\right)m_i,
\end{equation}
where $\rho$ is the Weyl vector of the residual group and thus equals  half the sum of the unbroken roots:
\begin{equation}
\rho = \frac{1}{2} \sum_{j=s+1}^r \alpha_j.
\end{equation}
In equation (\ref{eqn:dimnonabel}) one sums over the broken roots and thus also over the topological charges.The dimensions of the framed moduli spaces have two important properties. First for $g = g' + g''$ with topological charges $m_i = m'_i + m''_i$ we have 
\begin{equation}
\text{dim }\mathcal M (\hat r_0, \Phi_0, m_i)= \text{dim }\mathcal M (\hat r_0, \Phi_0, m'_i)+ \text{dim }\mathcal M (\hat r_0, \Phi_0, m''_i).
\end{equation}
Second if the residual gauge group equals the maximal torus $U(1)^r$ in $G$ so that there are no holomorphic charges the dimension formula above reduces to Weinberg's formula
\begin{equation}
\label{eqn:dimabel2}
\text{dim }\mathcal M (\hat r_0, \Phi_0, m_i)= 4\sum_{i=1}^r m_i.
\end{equation}
We thus see that equation (\ref{eqn:dimnonabel}) for the dimension of the framed moduli space is a generalization of Weinberg's result. More importantly we find that dimensions of the framed moduli spaces respect the addition of charges in the Murray cone.\\ 
\\
The dimensions of the framed moduli spaces are compatible with the addition of charges in the Murray cone. These dimensions do not depend on the holomorphic components. Naively it thus seems we can safely ignore these components. Nevertheless, from a physical perspective one is forced to take the holomorphic charge into account because it determines the allowed electric charge of a monopole as we shall discuss in a moment. It is thus very interesting to know how the holomorphic charges affects the fusion of single monopoles into multi-monopole configurations.\\
\\
If we want to take the holomorphic charges into account we should consider the strata within the framed moduli spaces. These strata were introduced in section \ref{subsection:fulframed}. For a given stratum $G_0$ is fixed up to the action of the residual gauge group and hence the holomorphic components of $g$ are given up to Weyl transformations. The dimensionality of the stratum corresponding to $g$ can be expressed in terms of the reduced magnetic charge as was shown by Murray and Singer \cite{Murray:2003mm}.\\
Let $g$ be any charge in the Murray cone and $\tilde g$ its reduced magnetic charge. Remember that $\tilde g$ is simply the lowest charge in the orbit of $g$ under the action of the residual Weyl group.  The reduced magnetic charge can thus be expressed as:
\begin{equation}
\label{eqn:redexp}
\tilde g = \sum_{i=1}^s m_i \alpha_i^* + \sum_{j=s+1}^r h_j \alpha_j^*.
\end{equation}
The dimensionality of the corresponding stratum $\mathcal M_g$ in the framed moduli space $\mathcal M(\hat r_0, \Phi_0, m_i)$ is given by:
\begin{equation}
\label{eqn:dimstrata}
\text{dim }\mathcal M_g = \sum_{i=1}^s 4m_i  + \sum_{j=s+1}^r 4h_j + \text{dim }C(\Phi_0) - \text{dim } C(\Phi_0)\cap C(G_0). 
\end{equation}
$C(\Phi_0) \in G$ is the centralizer subgroup of the Higgs VEV, i.e.~it is simply the residual gauge group $H$. Similarly, $C(G_0)\in G$ is the centralizer of the magnetic charge. Hence the fourth term in the equation above equals the dimensionality of the subgroup in $H$ that leaves $G_0$ invariant.  So the last two terms in equation (\ref{eqn:dimstrata}) express the dimension of the orbit of the magnetic charge $G_0$ under the action of the residual gauge group.\\
 In figure \ref{fig:stratadimsu2} we have worked out formula (\ref{eqn:dimstrata}) for $SU(3) \to U(2)$ for each charge in the Murray cone. In this particular case the $H$-orbits of the magnetic charges are either 2-spheres  or they are trivial.\\ 
\\
The next goal is to relate the dimensions of the strata to the generators of the Murray cone found in section \ref{subsect:genmurray}, the monopoles with unit topological charges. In the abelian case discussed previously  such a relation is obvious. Since there are no stratifications the moduli space dimensions are proportional to the topological charges. In the true non-abelian case such a simple relation is distorted by the centralizer terms in formula (\ref{eqn:dimstrata}).  This is easy to see in the $SU(3)$ example in figure  \ref{fig:stratadimsu2}. Therefore we shall have to leave  these centralizer terms out in our analysis. Since the centralizer terms correspond to the orbit of the magnetic charges under the action of residual gauge group, discarding the centralizer terms amounts to restricting to the fully framed moduli spaces introduced in section \ref{subsection:fulframed}.\\
\\
There are good arguments to discard the centralizer terms in the present discussion  or at least to treat them on a  different footing than the remaining terms in (\ref{eqn:dimstrata}). The centralizer terms count the dimensions of the orbit of the magnetic charge under the action of the electric group. Naively one would thus expect that this orbit is related to the electrical properties of the monopoles.  Such a picture is flawed because already at the classical level there is a topological obstruction for implementing the full residual electric group $H \subset G$ globally as has been proven by various authors \cite{Abouelsaood:1983gw,Balachandran:1983fg,Nelson:1983bu,Horvathy:1984yg,Horvathy:1985bp}. This obstruction is directly related to the fact that a magnetic monopole defines a nontrivial $H$-bundle on a sphere at infinity. A subgroup $H'\subset H \subset G$ is implementable as a global symmetry in the background of a monopole if the transition function (\ref{eqn:equator2})
\begin{equation}
\mathcal G(\varphi)= \exp\left(\frac{ie}{2\pi}G_0\varphi\right)
\end{equation} 
is homotopic to a loop in
\begin{equation}
Z_H(H')=\{ h\in H : hh'=h'h ~~\forall h' \in H\}
\end{equation}
the centralizer of $H'\subset H$. Note that the maximal torus $U(1)^r\subset H$ is always implementable. As as rule of thumb one finds that $H'$ can be non-abelian if up to unbroken coroots the magnetic charge has one or more vanishing  weights with respect to the non-abelian component of  $H$. This follows from the fact that the holomorphic components of the magnetic charge are not conserved under smooth deformations.\\
There is an even stronger condition on the electric symmetry that can be realized in the monopole background. One can show \cite{Balachandran:1983fg} that the action of the residual electric group maps finite energy configurations to monopole configurations with infinite energy if the magnetic charge is not invariant. The interpretation is that all BPS configurations with finite energy whose magnetic charges lie on the same electric orbit are separated by an infinite energy barrier.\\
Classically one thus finds that only if the generators of the residual gauge group H commute with the magnetic charge  one can define a global rigid action of H. In other words the monopole effectively breaks the symmetry further down so that only the centralizer group can be realized as a symmetry group. For example in the case that $SU(3)$ is broken to $U(2)$  monopoles with magnetic charge $g=2\alpha_2$ can only carry electric charges under $U(1)^2$, while monopoles in the same framed moduli space with $g=2\alpha_2 + \alpha_1$ might carry charges under the full residual $U(2)$ group. These obstructions persist at the semi-classical level \cite{Abouelsaood:1982dz,Abouelsaood:1983gw,Weinberg:1982ev}.\\
\\
The dimensions of the fully framed moduli spaces have a simple expression in terms of the topological and holomorphic components of the reduced magnetic charge $\tilde g$ \cite{Murray:2003mm}:
\begin{equation}
\label{eqn:dimfulfram}
\text{dim } \mathcal M(\mu,\Phi_0, G_0) = \sum_{i=1}^s 4m_i  + \sum_{j=s+1}^r 4h_j. 
\end{equation}
\begin{figure}[!ht]
\centerline{
\includegraphics{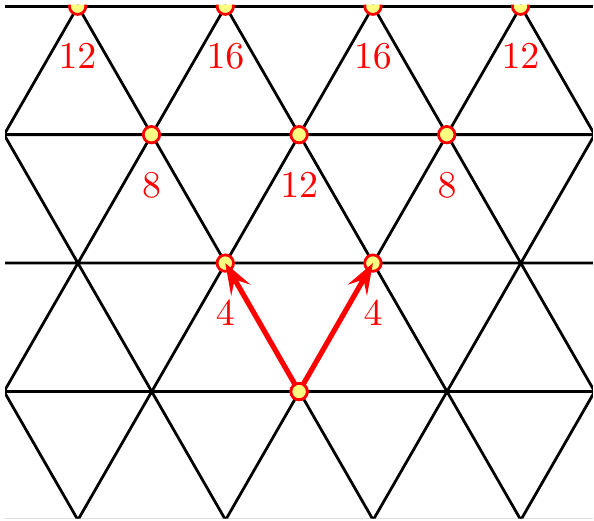}
}   
\caption{Dimensions for the fully framed  moduli spaces for $SU(3)$ broken to $U(2)$, and the generators of the Murray cone. The dimensions are only additive if one moves along the central axis of the cone or away from it.}
\label{fig:fullydimsu2}
\end{figure}
\begin{figure}[!ht]
\centerline{
\includegraphics{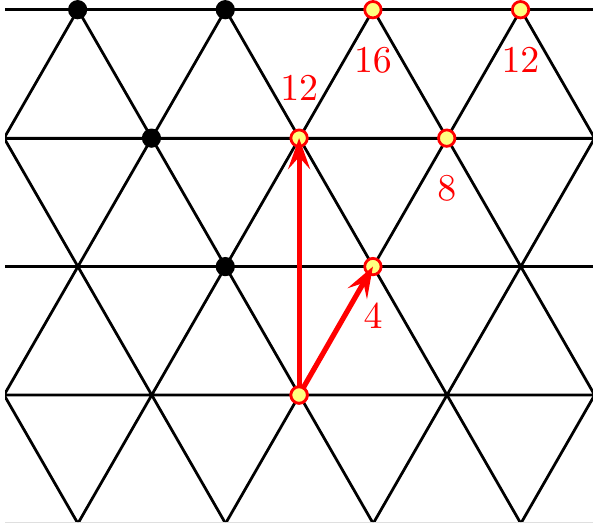}
}   
\caption{The fundamental Murray cone for $SU(3)$ broken to $U(2)$. In this example the magnetic charge lattice is interpreted as the weight lattice of $U(2)$. The fundamental Murray cone is the intersection of the full cone with the fundamental Weyl chamber of the $U(2)$ weight lattice.  The dimensions of the fully framed moduli spaces are additive under the composition of the generators depicted by the arrows.}
\label{fig:fundcone}
\end{figure}
\\
Previously we have found that the Murray cone is spanned by the magnetic charges with unit topological charges. We might hope that the dimensions of the fully framed moduli spaces behave additively with respect to the expansion into these indecomposable charges as was the case for the framed moduli space. Such additive behaviour does indeed occur, but only partially. For instance the case of $SU(3) \to U(2)$ is worked out in figure \ref{fig:fullydimsu2}. The additivity of the moduli space dimensions still holds as long as we stick to one of the Weyl chambers of the cone, defined with respect to the residual Weyl action.\\
Apparently the dimensions of the fully framed moduli spaces are not compatible with the Murray cone in general. However, as we will prove below these dimensions are compatible with the fundamental Murray cone.\\
In the abelian case this is obviously true. The Weyl group of the residual group is now trivial and there is no additional identification within the cone. Therefore we can refer back to the previous sections where we found that the generating charges have unit topological charge and that the dimensionality of the moduli space is proportional to the total topological charge. Our favourite example in the truly non-abelian case $SU(3)\to U(2)$ is worked out in figure \ref{fig:fundcone}. The generators of the fundamental Murray cone are easily recognized and the additivity of dimensions is easily confirmed.\\
We claim that the additivity of the moduli space dimensions with respect to a decomposition in generating charges of the fundamental Murray cone holds in general. Without an explicit set of generators it seems we cannot prove this directly. However, it suffices to check the additivity for every pair of charges in the fundamental cone.
\begin{prop}
\label{prop:add}
For any pair of magnetic charges $g$ and $g'$ in the fundamental Murray cone we have for the fully framed moduli spaces $dim\,\mathcal M_g + dim\,\mathcal M_{g'} = dim\, \mathcal M_{g+g'}$ .
\end{prop}
\proof Recall from equation (\ref{eqn:dimfulfram}) that the dimensions of the fully framed moduli space are proportional to the topological and holomorphic charges of the reduced magnetic charge. We thus have to show that the topological and holomorphic charges add. These charges are given by the inner product of the reduced magnetic charge with respectively the broken and unbroken fundamental weights as explained in section \ref{subsect:quantBPS} and \ref{subsect:murraycond}. For example $m_i=\lambda_i \cdot \tilde g$. Next we note that there exists a Weyl transformation $w\in W(H)\subset W(G)$ that maps the fundamental Weyl chamber to the anti-fundamental Weyl chamber. Thus if $g, g', g''$ lie in the fundamental Murray cone and $g''=g+g'$ the reduced magnetic charges satisfy $\tilde g'' = w(g'') = w(g) + w(g') = \tilde g + \tilde g'$. As a last step we find that $m_i'' =   \lambda_i \cdot \tilde g '' = \lambda_i \cdot (\tilde g + \tilde g') = m_i + m_i'$. Similar results hold for the holomorphic charges.\endproof 
The dimensions of the fully framed moduli spaces  only respect the addition of charges in the Murray cone if the charges are restricted to one Weyl chamber, for example the fundamental Weyl chamber. This is consistent with our conclusion at the end of section \ref{subsect:murraycond} that the magnetic charge sectors are labelled by weights in the fundamental Weyl chamber of the residual dual group.\\ 
\\
In this section we have established a non-abelian generalization of Weinberg's analysis for abelian monopoles: we have shown that the dimensions of the fully framed moduli spaces respect the addition of magnetic charges within the fundamental Murray cone. Just as Weinberg we are now led to the conclusion that there is a distinguished set of basic monopoles. The charges of these basic monopoles correspond to the generators of the fundamental Murray cone. The remaining charges in the fundamental Murray cone are then associated with multi-monopole solutions.\\
For maximal symmetry breaking the set of basic monopoles coincides with the monopoles with unit topological charge. In our proposal this is not true in the general case. There can be basic monopoles with non-minimal topological charges. In the next section we shall discuss additional evidence to support our conclusion that basic monopoles are always indecomposable, even if they have non-minimal topological charges.  
%
%
%
\section{Fusion properties of non-abelian monopoles}
\label{sect:fusionprop}
In the previous sections we argued that smooth BPS mono\-poles with non-trivial charges can consistently be viewed as multi-monopole solutions built out of BPS configurations with minimal charges. These classical fusion rules cannot always be verified directly because of the complexity of the BPS equations. 
In this paper we have therefore gathered all available circumstantial evidence. These consistency checks can be organized into four different themes: the existence of generating charges and the consistent counting of moduli space parameters have been discussed in the previous sections. Below in section \ref{sect:splitjoin} and \ref{sect:clouds} we shall study some examples where one can verify the classical fusion rules directly. For singular BPS monopoles there is a similar  set of generating charges, a consistent counting of parameters and a consistent way to patch  classical solutions together as we discuss in section \ref{splitjoinsing}.  These analogies form a remarkable hint suggesting that the classical fusion rules we have found for smooth BPS monopoles are indeed correct. Finally in section  \ref{sect:discussion} we look ahead and discuss how this analogy between singular and smooth BPS monopoles might help us to derive the semi-classical fusion rules of smooth BPS monopoles and conversely how to get a better understanding of the generalized electric-magnetic fusion rules in the singular case.  
\subsection{Patching smooth BPS solutions}
\label{sect:splitjoin}
The first hint revealing the existence of multi-monopole solutions built out of certain minimal monopoles comes from the fact that there is a small set of indecomposable charges generating the full set of magnetic charges. In this section we use results of Taubes obtained in \cite{Taubes:1981gw} to show that certain monopoles with non-trivial charges are indeed multi-monopole solutions respecting the decomposition of the magnetic charge into generating charges. We shall first discuss maximal symmetry breaking. In this case all monopoles with higher topological charges are manifestly seen to be multi-monopoles. Second we shall deal with non-abelian residual gauge groups. In this case Taubes' result gives a consistency check for the classical fusion rules.\\
\\
For maximal symmetry breaking the set of magnetic charges corresponds to the Murray cone  and is generated by the broken simple coroots. For each of these coroots an exact solution is known. These are  spherically symmetric $SU(2)$ monopoles \cite{Bais:1978fw,Weinberg:1979zt,Weinberg:1982ev}. For $G$ equal to $SU(2)$ one has the usual 't Hooft-Polyakov monopole \cite{'tHooft:1974qc,Polyakov:1974ek}, while for higher rank gauge groups one can embed 't Hooft-Polyakov monopoles via the broken simple roots. Since these monopoles have unit topological charges they are manifestly indecomposable.\\
Exact solutions are also known in other cases. It was shown by Taubes \cite{Taubes:1981gw} that there are solutions to the BPS equation for any charge $g= m_i\alpha_i^*$ with $m_i\!>\!0$. Hence for all charges in the Murray cone solutions exist.  These solutions are constructed out of superpositions of embedded $SU(2)$ monopoles. The constituents are chosen such that the sum of the individual charges matches the total charge. These solutions become smooth solutions of the BPS equations if the constituents are sufficiently separated. This proves that for all magnetic charges with higher topological charges multi-monopole solutions exist.\\
One might wonder if all solutions with higher topological charges are indeed multi-monopole solutions. For any given topological charge the framed moduli space is connected. Thus any point in this moduli space is connected to another point corresponding to a widely separated superposition as described by Taubes. Any monopole configuration can thus be smoothly deformed so that the individual components are manifest. This does indeed show that any smooth abelian BPS monopole can consistently be viewed as multi-monopole configuration built out of indecomposable monopoles.\\
\\
The multi-monopole picture above for maximal symmetry breaking  can be generalized to arbitrary symmetry breaking. For any given topological charge there exist smooth solutions of widely separated monopoles.  According to Taubes the building blocks of these smooth solutions correspond to the $SU(2)$ monopoles embedded via the broken simple roots. The framed moduli space does not depend on the full magnetic charge, but only on the topological components. Moreover the framed moduli space is always connected. We now find that any solution of the BPS equation with higher topological charges can be deformed to a configuration which is manifestly a multi-monopole solution.\\
However, this decomposition via widely separated multi-monopole solutions does not respect the additive structure of the Murray cone unless we completely ignore the holomorphic charges. In the previous section we argued that this does not make sense from a physical perspective because the allowed electric excitations depend on the holomorphic charges. Moreover we have found that if we take these holomorphic charges into account we should restrict the magnetic charges to lie in the fundamental Murray cone. The appropriate moduli spaces to consider in this situation are the fully framed moduli spaces. The question now is if these fully framed moduli spaces contain configurations which can be interpreted as widely separated monopoles.\\
With the results of Taubes we can answer this question unambiguously for one of the fully framed moduli spaces in the set of spaces defined by the topological charge. We start out with a magnetic charge that is equal to a sum of unbroken simple coroots. Such a charge does not lie in the fundamental Murray cone, but instead in the anti-fundamental Weyl chamber of the Murray cone. This implies that there is a Weyl transformation that maps such a magnetic charge to the fundamental Murray cone. Similarly, there is a related large gauge transformation that maps  Taubes' multi-monopole configurations to new solutions of the BPS equation. These transformed configurations are again widely separated superpositions, but now the buildings blocks correspond to $SU(2)$ monopoles embedded via the Weyl transformed broken roots. Note that magnetic charges of these constituents are precisely the generators of the fundamental Murray cone with unit topological charges.\\
We thus obtain the following result: let $g$  equal a sum of generators of the fundamental Murray cone with unit topological charges. The fully framed moduli space corresponding to $g$ has a subset of configurations that are manifestly multi-monopole solutions. Since the fully framed moduli space is connected any monopole with charge $g$ can be interpreted as a multi-monopole solution.\\
\\
The considerations above only involved indecomposable charges corresponding to simple coroots. For a maximally broken gauge group this is sufficient to provide convincing evidence for the multi-monopole picture. In the non-abelian case one should also take other generators into account. We will come back to this in the next section.
\subsection{Murray cone vs fundamental Murray cone}
\label{sect:clouds}
One problem  we encounter in this paper  is that it is not completely clear what the full set of magnetic charges is supposed to be, either the Murray cone or the fundamental Murray cone. By the same token it is a priori not clear what the truly indecomposable monopoles are, the fundamental monopoles or the basic monopoles. The fundamental Murray cone is slightly favoured because because the large gauge transformations have been modded out. On the other hand not all generating charges of the fundamental Murray cone have unit topological charges, while the fundamental monopoles generating the Murray cone do. This suggest that the monopoles corresponding to the generators of the fundamental Murray cone, the basic monopoles might be decomposable into fundamental monopoles related to the generators of the Murray cone. What seems to settle this issue though is that there is only a consistent counting of moduli space parameters if we restrict to the fundamental Murray cone. The indecomposability for the basic monopoles with non-minimal topological charges can be understood from the existence of so-called cloud parameters. These clouds emerge as soon as one attempts to split a basic monopole into fundamental monopoles. Below we explain this for the case that $G=SU(3)$ is broken to $U(2)$.\\
\\
The case $SU(3)\to U(2)$ is an interesting example to discuss issues regarding composition and decomposition because some of the corresponding non-trivial moduli spaces  have been thoroughly investigated. Specifically for $g=2\alpha_2 + \alpha_1$, one of the generators of the fundamental Murray cone, the 12 dimensional fully framed moduli space  and its metric  have been found Dancer \cite{Dancer:1992kn}. Determining the isometries of the metric reveals the nature of almost all of the 12 parameters. 3 parameters are related to $\mathbb R^3$, the center of mass position in space. The action of  $U(2)\times SO(3)$ shows the presence of 3 rotational degrees of freedom and 4 large gauge modes. After the removal of translations, gauge freedom and rotations one is left with a 2-dimensional space. This space turns out to be parameterized by  $k$ and $D$ with $0\leq k \leq 1$ and $0\leq D < \tiny{\frac{2}{3}}K(k)$, where $K(k)$ denotes the first complete elliptic integral $K(k)\!=\!\small{\int_0^{\frac{\pi}{2}}(1-k^2\sin^2(\theta))^{-1/2}d\theta}$ \cite{Dancer:1992kn}. The interpretation of these two parameters seems somewhat mysterious. To understand their significance, the behaviour of the BPS solutions have been studied numerically for various values of these parameters \cite{Dancer:1992kj,Dancer:1997zx,Irwin:1997ew}. See section 8 of \cite{Bais:1997qy} and section III.B of \cite{Houghton:2002bz} for a review.\\
There is a subset of solutions where the energy density has two maxima symmetrically positioned about the center of mass. If the parameter $D$ is increased the peaks of the energy density becomes more pronounced and move further from the center of mass. This seems to indicate that certain solutions can be viewed as a pair of widely separated particles. The question that comes to mind now is the following: do these widely spaced lumps correspond to a pair of monopoles with unit topological charge? In this particular case there is only one broken simple root, and hence there is only one class of embedded $SU(2)$ monopoles with unit topological charge giving rise to a  4 dimensional fully framed moduli space. Taubes has shown that such solutions can be patched together yielding widely separated solutions with higher topological charges \cite{Taubes:1981gw}. A pair of these patched solutions with magnetic charge $\tilde g=\alpha_2 +\alpha_1$ would give a configuration with total charge $2\alpha_2 + 2\alpha_1$. We thus see that these solutions do not lie in the 12 dimensional Dancer moduli space but in the neighbouring 8 dimensional fully framed moduli space. Before drawing any conclusions we recall one subtle point. All monopoles of equal topological charge lie in one connected moduli space, which is divided up into strata. By dividing out large gauge transformations these strata reduce to the fully framed moduli spaces we discussed here. To be more precise the 8 dimensional fully framed moduli space related to $g=2\alpha_2 +2\alpha_1$ lies in a 10 dimensional stratum which is the boundary of the  Dancer moduli space. A measure for the distance to the boundary of the Dancer moduli space is given by $1/a$ where $a=D/(K(k)-\frac{3}{2}D)$. The widely separated monopoles in the Dancer moduli space, and as a matter of fact any configuration in the Dancer moduli space can be deformed into widely separated monopoles discussed by Taubes. As we explained in section \ref{sect:splitjoin} this decomposition does not respect the additive structure of the Murray cone nor the addition of charges in the fundamental Murray cone. What is even more striking is that this deformation will give rise to highly non-localized degrees of freedom in the form of a non-abelian cloud.\\
\\
A rather insightful computation to illustrate the appearance of the non-abelian cloud as one moves to the boundary of the Dancer moduli space is discussed by Irwin \cite{Irwin:1997ew}. In his paper Irwin computes the asymptotic behaviour for the magnetic field of axially symmetric trigonometric monopoles ($k=0$) in the Dancer moduli space as $a\to \infty$. In the string gauge the asymptotic fields in this limit are given by:
\begin{equation}
\begin{split}
\label{eqn:cloudsolution}
\Phi =\,&\Phi_0 - \frac{1}{er}t_0 - \frac{1}{er(1+r/a)}t_3  \\ 
*F =\,&\frac{1}{er^2}t_0 dr + \frac{1+2r/a}{er^2(1+ r/a)^2} t_3 dr +\frac{1}{ear^2(1+r/a)^2}\left(t_1d\theta - \sin \theta t_2 d\varphi\right) \\ 
A =\,& -\frac{1}{e}\cos\theta \left(t_0 + t_3\right) d\varphi - \frac{1}{e(1+a/r)}\left(t_2d\theta + \sin\theta t_1d\varphi\right).
\end{split}
\end{equation}
The expectation value $\Phi_0$ is proportional to $t_0$. The matrices $t_i$ are the generators of the residual gauge group $U(2)\subset SU(3)$: $t_0 = (\alpha_1+2\alpha_2)\cdot H =\sqrt{3}H_2 $, $t_1 = \frac{1}{2}\left( E_{\alpha_1} + E_{-\alpha_1}\right)$, $t_2= -\frac{i}{2}\left(E_{\alpha_1} - E_{-\alpha_1}\right)$ and  $t_3= \alpha_1\cdot H = H_1$. With these conventions the commutation relations for the $SU(2)$ generators are given by $[t_i,t_j]= i\epsilon_{ijk}t_k$. Obviously we also have $[t_0,t_i]=0$. 
As a side remark we  note that (\ref{eqn:cloudsolution}) gives an exact solution of the BPS equations for $U(2)$ which is singular at r=0.\\
The behaviour of the solution above shows that the non-abelian fields penetrate outside the core of the monopoles up to some finite distance determined by the parameter $a$. Beyond this distance  the magnetic field becomes abelian, which is in agreement with the boundary condition at infinity. The interpretation of this observation is that the monopoles are surrounded by a non-abelian cloud screening the non-abelian charge. As one moves all the way to the boundary of the Dancer moduli space and $a$ becomes infinite the cloud gets diluted so that the non-abelian field yields to infinity resulting in non-vanishing holomorphic charges.\\
\\
This whole exposition does lead us to an important conclusion: the behaviour of the Dancer monopoles shows us that these configurations can indeed be split up into separate lumps. At the same time this separation yields a non-localized degree of freedom. Since this cloud parameter does not correspond to one monopole or the other, the two widely spaced lumps are not the same as they would be on their own. Therefore the Dancer monopoles cannot be decomposed. In this particular example we thus see that basic monopoles are indeed indecomposable. We expect that similar arguments should hold in general.
\subsection{Patching singular BPS solutions}
\label{splitjoinsing}
There are striking similarities between the results obtained in this paper for smooth BPS monopoles and results obtained by Kapustin and Witten regarding singular BPS monopoles \cite{Kapustin:2006pk}. In the context of singular monopoles we shall discuss the existence of  fundamental and basic monopoles, consistent counting of moduli space parameters,  patching of classical solutions and the indecomposability of basic monopoles with non-trivial topological charges.\\
\\   
The magnetic charge lattice for singular monopoles in a gauge theory with gauge group $H$ is determined by the Dirac quantization condition. As we reviewed in section \ref{sect:singcondition} this lattice can be identified with the weight lattice $\Lambda(H^*)$ of the GNO or Langlands dual group.  The magnetic weight lattice contains an important subset, the set of magnetic charge sectors, which is obtained by modding out the Weyl group of $H^*$. This subset can thus be identified with the fundamental Weyl chamber in $\Lambda(H^*)$. Modding out the Weyl group of $H^*$ is natural because a magnetic charge $\lambda \in \Lambda(H^*)$ is only defined up to Weyl transformations. Note that these Weyl transformations acts as large gauge transformations on the BPS solutions.\\ 
\\ 
The existence of generating charges within the weight lattice $\Lambda(H^*)$ of the dual gauge group  and within its fundamental Weyl chamber has been discussed in sections \ref{section:genweight} and \ref{subsect:genfundWeyl}. These generating charges correspond to what we define as respectively fundamental monopoles and basic monopoles.  The basic monopoles, not the fundamental monopoles, form the building blocks of singular multi-monopole solutions of the BPS equations just as we concluded for smooth BPS solutions. This is seen indirectly by analyzing the moduli space parameters.\\
\\ 
The moduli spaces for singular BPS monopoles introduced by Kapustin and Witten are  spaces of so-called Hecke modifications and correspond to orbits in the affine Grassmannian. For further details we refer to \cite{Kapustin:2006pk} and reference therein. It is important that these moduli spaces are labelled by a dominant integral weight in the weight lattice of the dual gauge group $H^*$. We also note that these moduli spaces are closed under large gauge transformations, hence magnetic charges on one Weyl orbit correspond to the same moduli space. For completeness we mention that the compactifications of these moduli spaces are singular. The singular subspaces in $\overline {\mathcal M}_\lambda$ correspond to the moduli spaces $\mathcal M_{\lambda'}$ where $\lambda' <\lambda$.\\
\\
The dimensionality of an orbit $\mathcal M_\lambda$ in the affine Grassmannian labelled by a dominant integral weight $\lambda \in \Lambda(H^*)$ is given by \cite{Lusztig}:
\begin{equation}
\text{dim}\: \mathcal M_\lambda = 2\lambda\cdot\rho,
\end{equation}
where $\rho$ is the Weyl vector of $H$ and thus equals half the sum of the simple roots of $H$, see e.g.~\cite{braverman-1999} for a brief summary. We now immediately find for a pair of dominant integral weights $\lambda$ and $\lambda'$:
\begin{equation}
\text{dim}\: \mathcal M_\lambda +  \text{dim}\: \mathcal M_{\lambda'} = \text{dim} \:\mathcal M_{\lambda + \lambda'}.
\end{equation}
Here we use the fact that sum of two dominant integral weights is again a dominant integral weight. We thus see that the moduli space dimensions respect the addition of charges in the fundamental Weyl chamber of $H^*$. It is not difficult to see that these dimensions are not consistent with the addition of charges in the complete weight lattice of $H^*$. Similar results where obtained in section \ref{sect:stratadims} for the Murray cone and the fundamental Murray cone.\\
\\
The formalism in which Kapustin and Witten work is so powerful that one can quite explicitly see that all singular monopoles with non-basic charges are indeed multi-monopole solutions. This is related to the fact that the singularities of the compactified moduli spaces can be blown-up in a very specific way. We briefly sketch how this works. Singular BPS monopoles correspond to 't Hooft operators which create the flux of a Dirac monopole with a singularity at  a point $p\in\mathbb R^3$. These 't Hooft operators can in turn be identified with Hecke operators. The Hecke operators act on vector bundles over  $\mathbb C $ in this case in such a way that the trivialization outside a preferred point on  $\mathbb{C}$ is respected. To achieve this relation $\mathbb R^3$ is identified with $\mathbb C \times \mathbb R$. For a non-zero magnetic charge the Hecke operator maps a trivial bundle to a non-trivial bundle. These two bundles over $\mathbb C$ are identified with pullback bundles of the nontrivial bundle over $\mathbb R^3\setminus \{p\}$ corresponding to the singular BPS configuration. The two embeddings of $\mathbb C$ are chosen at opposite sides of $p \in \mathbb C \times \mathbb R$. Note that the isomorphism class of the resulting modified bundle does not only depend on the magnetic charge but also in a certain way on the trivialization of the trivial bundle one started out with. This why one Hecke operator gives rise to a space of Hecke modifications.\\
A relevant but actually not very deep observation is that all Hecke operators can be decomposed as a sequence of basic Hecke operators and thus all 't Hooft operators can be decomposed as sequence of basic 't Hooft operators. These basic operators create the flux of a Dirac monopole associated to a basic charge in the fundamental Weyl chamber of $H^*$. An important as well as deep consequence of the identification of 't Hooft operators and Hecke operators is that the resulting sequence of basic 't Hooft operators can be separated in space. Each basic 't Hooft operator is positioned between two copies of $\mathbb C \subset \mathbb R^3$ and the associated Hecke operators map one bundle over $\mathbb C$ to the other bundle at the reverse side of the singularity. The resulting  bundles over $\mathbb C$ can be considered as as series op pullback bundles in a bundle over $\mathbb R^3$ corresponding to a series of smoothly patched BPS solutions.\\
\\
Singular BPS solutions corresponding to basic monopoles may have non-trivial topological charges just as we have seen for smooth BPS solutions. One might  again wonder if such basic monopoles can be split up into fundamental monopoles which do have unit topological charges. Intuitively this does not seem difficult. One would expect that there exists an exact multi-monopole solution of widely separated fundamental monopoles in the same topological sector. Because all  spaces of Hecke modifications in one topological sector are connected one can now deform the original monopole into a manifest multi-monopole solution. Just as for smooth monopoles this deformation does not respect the holomorphic charges. There is also a more subtle way to look at this holomorphic obstruction.\\
As an example we consider $H^*=U(2)$. This case has been worked out in quite some detail by Kapustin and Witten. The basic monopole with unit topological charge corresponds to the highest weight $\lambda$ of the fundamental representation of $U(2)$. The basic monopole with topological charge equal to 2 has a magnetic charge given by $2\lambda-\alpha$, where $\alpha$ is the simple root of $SU(2)$. The compactification of $\mathcal M_{2\lambda}$ is given by the singular space $\mathbb{WCP}_2(1,1,2)$. The singularity corresponds to the $0$-dimensional space $\mathcal M_{2\lambda-\alpha}$. The singularity of  $\mathbb{WCP}_2(1,1,2)$ can be blown-up to obtain $\mathbb{CP}_1\times \mathbb{CP}_1$. Since $\mathbb{CP}_1= \mathcal M_{\lambda}$ the blow-up obviously gives the moduli space of two separated fundamental monopoles. On thus sees that the a basic monopole with topological charge 2  can be deformed into two separate monopoles only if one attributes extra degrees of freedom by embedding the moduli space as a singularity in a larger space and only if these degrees of freedom are changed in a nontrivial way by blowing-up the singularity. It follows for $U(2)$ that classically basic monopoles are not truly separable into fundamental monopoles. For general monopoles similar arguments should
 hold.\\
Note that for smooth monopoles we used the emergence of non-localized degrees of freedom to show that the basic $U(2)$-monopoles are indeed indecomposable. Though the motivation via clouds is quite different from the argument used right above in the singular case the result is the same: the charges of the indecomposable monopoles in the smooth theory are subset of the indecomposable charges for singular monopoles. In that sense the classical fusion rules for smooth BPS monopoles are consistent with those for singular BPS monopoles.  
\subsection{Towards semi-classical fusion rules}
\label{sect:discussion}
In the literature it has often been assumed that the BPS solutions corresponding to weights of the  fundamental representation of $H^*$ give rise to a single $H^*$-multiplet \cite{Bais:1978fw,Goddard:1981kf,Goddard:1981rv,Lee:1996vz,Dorey:1995me,Bais:1997qy,Schroers:1998pg} as would be favourable to the conjecture that these monopoles can be regarded as massive gauge particles of the dual theory. This proposal runs into trouble in particularly for non-simply laced gauge groups  because the electric action on the classical solutions does clearly not commute with the magnetic action of the residual dual group on the magnetic charges. From the classical fusion rules we find that smooth BPS solutions, and actually also singular BPS solutions, are labelled by dominant integral weights in the weight lattice of the residual dual group $H^*$. This suggest that each electric orbit in the magnetic charge lattice of $H^*$ thus gives rise to a unique $H^*$-multiplet. This form of the  GNO-duality conjecture has been proven by Kapustin and Witten in the case of singular BPS monopoles \cite{Kapustin:2006pk}. They show that the semi-classical fusion rules for singular BPS monopoles are indeed the fusion rules of $H^*$. Since the classical fusion rules for singular and smooth  BPS monopoles are completely analogous and because the semi-classical fusion rules must also agree one can expect that a similar approach can be used to derive the semi-classical fusion rules in the smooth case. It is not immediately clear though how such a program can be realized and some major hurdles have to be overcome. We shall discuss this shortly.\\
\\
In the Kapustin-Witten approach the semi-classical fusion rules are found from the quantum mechanics on the moduli spaces. A similar strategy but with a less ambitious goal in mind was adopted in the case of smooth monopoles by Dorey et al. in \cite{Dorey:1995me}. These authors tried to give a consistent counting of states, an attempt  that turned out not to be completely successful. In hindsight we can understand that the problem was caused by the fact Dorey et al.~did not use the same moduli spaces as Kapustin and Witten. The moduli spaces used by Kapustin and Witten can be identified with orbits in the affine Grassmannian labelled by the magnetic charges in the weight lattice of $H^*$. These orbits do contain the orbits of the magnetic charge in $\mathfrak h$, the Lie algebra of $H$, under the action of the  gauge group $H$ as used in \cite{Dorey:1995me}.  Only if the magnetic charge labels a so-called minuscule representation the orbit in the affine Grassmannian is isomorphic to the orbit in the Lie algebra of $H$. In these cases the number of ground states of the quantum mechanics on the orbit in the Lie algebra agrees with the dimension of the irreducible representation labelled by the magnetic charge. In other cases the orbit in $\mathfrak h$ is a non-trivial subspace within the orbit in the affine Grassmannian. The degeneracy of the ground state of the quantum mechanics on the orbit in $\mathfrak h$  under-estimates the dimension of the magnetic representations.\\
\\
If one wants to retrieve a counting of states consistent with the irreducible representations of $H^*$ as well as the fusion rules of $H^*$ one is forced to consider the orbits in the affine Grassmannian. The problem is that it is only partially clear how these magnetic moduli spaces are to appear within the full moduli spaces of smooth BPS monopoles.  What is consistent though is that the orbits of the magnetic charges under the electric action, which are part of the related orbits in the affine Grassmannian, have to be treated on a different footing within  the framed moduli spaces as we discussed in section \ref{sect:stratadims}.\\
\\
One of the drawbacks of the approach used by Kapustin and Witten is that it is not clear how to deal with electric excitations of magnetic monopoles even though one can introduce general Wilson-'t Hooft operators in for example certain $\mathcal N=2$ gauge theories \cite{Kapustin:2006hi}. If one manages for smooth monopoles to introduce the appropriate magnetic moduli spaces in combination with the fully framed moduli spaces then one obtains an interesting model to study electric-magnetic symmetry. This would be an important achievement because it is not known what this unified electric-magnetic symmetry is. It is clear though that this symmetry is not the group  $H\times H^*$ as originally proposed in \cite{Goddard:1976qe} because the magnetic charge effectively breaks the electric group as we discussed in section \ref{sect:stratadims}. In \cite{Schroers:1998pg} a unified electric-magnetic group was introduced for $H=U(n)$ which does respect this interaction between electric representations and magnetic charges. We plan to elaborate on this elsewhere \cite{kampmeijer}. 
\addcontentsline{toc}{section}{Acknowledgements} 
\section*{Acknowledgements}
We are grateful to the Kavli Institute for Theoretical Physics for hospitality during a visit in 2006. LK thanks Ilies Messamah, Lotte Hollands, Tom Koornwinder and Jochen Heinloth for helpful discussions. This work is part of the research programme of the 'Stichting voor Fundamenteel Onderzoek der Materie (FOM)', which is financially supported by the 'Nederlandse Organisatie voor Wetenschappelijk Onderzoek (NWO)'.
\appendix
\addcontentsline{toc}{section}{Appendix: The algebra underlying the Murray cone}
\section*{Appendix: The algebra underlying the Murray cone}
\label{appnd:reducgroup}
As announced in section \ref{subsect:genfundcone} we shall construct an algebraic object whose set of irreducible representations corresponds to the fundamental Murray cone. We do this in such a way that the fusion rules respect the fusion rules of the residual dual group $H^*$. We shall start out from what is roughly speaking the group algebra of $H^*$. Next we introduce its dual $F(H^*)$. By dualizing again we find an object $F^*(H^*)$  which again should be thought of as the group algebra of $H^*$. The difference however is that in this new form the group algebra can explicitly be truncated to $F^*_+(H^*)$ in such a way that the irreducible representations are automatically restricted to the fundamental Murray cone. The nice feature of our construction is that it is very general. Starting out from any Lie group and any subset of irreducible representations closed under fusion we can construct a bi-algebra which has a full set of irreducible representations corresponding to the subset one started out with and whose fusion rules match those of the group one started out with. At the end we briefly discuss the group-like object $H^*_+$ which has the same irreducible representations and the same fusion rules as $F^*_+(H^*)$. For most common consistent truncations of the weight lattice of $H^*$ one knows that $H^*_+$ is obtained from $H^*$ by modding out a finite group. If one restricts the weight lattice to the Murray cone however $H^*_+$ is not a group any more. \\
\\
As a group $H^*$ has a natural product and coproduct:
\begin{eqnarray}
h_1\times h_2 = h_1h_2 \\
\Delta(h)= h\otimes h.
\end{eqnarray}
In addition there is a natural unit $1$, co-unit $\epsilon$ and antipode $S$ by
\begin{gather}
1 = e \in H^* \\
\epsilon: h \mapsto 1 \in \mathbb C \\
S: h \mapsto h^{-1}. 
\end{gather}
For a finite group one can immediately define the linear extensions of these maps on the group algebra of $H^*$.  For a continuous groups there are several ways to define a vector space with this Hopf algebra structure. We shall circumvent this discussion by considering another algebra which is manifestly seen to be a vector space. This is the Hopf algebra corresponding to the matrix entries of the irreducible representations of $H^*$. Let $\pi^\lambda$ be such a representation. For the matrix entries we have:
\begin{equation}
\pi^\lambda_{mm'}: h\in H^* \to (\pi^\lambda(h))_{mm'} \in \mathbb C.
\end{equation}
The set of finite linear combinations of such maps is obviously a vector space. The resulting set turns out to be a Hopf algebra and inherits a natural product, coproduct, co-unit and antipode from $H^*$.
\\
The product in $F(H^*)$   is directly related to the product of representations and can thus be expressed in terms of Clebsch-Gordan coefficients, see for example chapter 3 of \cite{Biedenharn:1981} for the SU(2) case. The coproduct is much simpler because it merely reflects the fact that the product of $H^*$ is respected by the representations.
\begin{eqnarray}
\label{eqn:dualprod}
\pi^{\lambda_1}_{m_1 m_1'} \times \pi^{\lambda_2}_{m_2 m_2'} (h) = \pi^{\lambda_1}_{m_1 m_1'}  \otimes \pi^{\lambda_2}_{m_2 m_2'} (\Delta (h)) = \nonumber \\
\pi^{\lambda_1}_{m_1 m_1'} \otimes \pi^{\lambda_2} _{m_2 m_2'} (h\otimes h) = \pi^{\lambda_1}_{m_1 m_1'}(h)\pi^{\lambda_2}_{m_2 m_2'}(h)=\\
\sum_\lambda C^{\lambda_1 \lambda_2 \lambda}_{m_1, m_2, m_1 + m_2} C^{\lambda_1 \lambda_2 \lambda}_{m_1', m_2', m_1' + m_2'}\pi^\lambda_{m_1+m_2, m_1'+m_2'}(h) \nonumber
\end{eqnarray}
\begin{equation}
\begin{split}
\Delta(\pi^\lambda_{mm'})(h_1\otimes h_2) &= \pi^\lambda_{mm'}(h_1 \times h_2)= \\
\pi^\lambda_{mm'}(h_1 h_2)= \sum_s \pi^\lambda_{ms}(h_1)\pi^\lambda_{sm'}(h_2)&= \sum_s \pi^\lambda_{ms}\otimes \pi^\lambda_{sm'} (h_1\otimes h_2)
\end{split}
\end{equation}
The product and coproduct on $F(H^*)$ are thus completely defined by:
\begin{eqnarray}
\pi^{\lambda_1}_{m_1 m_1'} \times \pi^{\lambda_2}_{m_2 m_2'}\!\! &=& \!\! \sum_\lambda C^{\lambda_1 \lambda_2 \lambda}_{m_1, m_2, m_1 + m_2} C^{\lambda_1 \lambda_2 \lambda}_{m_1', m_2', m_1' + m_2'}\pi^\lambda_{m_1+m_2, m_1'+m_2'}  \\
\Delta(\pi^\lambda_{mm'}) &=& \sum_s \pi^\lambda_{ms}\otimes \pi^\lambda_{sm'}.
\end{eqnarray}
\\
To find the unit $1\in F(H^*)$ and the co-unit of $F(H^*)$ we note that these are defined in terms of their dual counterparts by:
\begin{gather}
1(h)= \epsilon(h)= 1   \in \mathbb C \\
\epsilon(\pi^\lambda_{mm'}) = \pi^\lambda_{mm'}(e)=\delta_{mm'} \in \mathbb C.
\end{gather}
From the first equation it follows that the unit in $F(H^*)$ is given by the  matrix entry  of the trivial irreducible representation of $H^*$ while the co-unit of $F(H^*)$ is related to the entries of the unit matrix in the $\lambda$-representation of $H^*$.\\
The antipode of $F(H^*)$ is defined by $S(\pi^\lambda_{mm'})(h)=\pi^\lambda_{mm'}(h^{-1})$. In the end however we will only be interested in the bi-algebra structure. To avoid unnecessary complication we will ignore the antipode.\\
\\
To retrieve the group algebra of $H^*$ we shall again take the dual $F^*(H^*)$ of $F(H^*)$. This space of linear functionals is generated by the basis elements $f_\mu^{ll'}$. These are defined in the standard way by:
\begin{equation}
f_\mu^{ll'}: \pi^\lambda_{mm'} \in F(H^*) \mapsto f_\mu^{ll'}(\pi^\lambda_{mm'})=\delta_{\mu\lambda}\delta_{lm}\delta_{l'm'} \in \mathbb C.
\end{equation}
The product and coproduct of $F^*(H^*)$ can be defined in terms of their counterparts in $F(H^*)$.
\begin{equation}
\begin{split}
&f_{\mu_1}^{l_1 l_1'} \times f_{\mu_2}^{l_2 l_2'} (\pi^\lambda_{mm'}) = f_{\mu_1}^{l_1 l_1'} \otimes f_{\mu_2}^{l_2 l_2'} (\Delta(\pi^\lambda_{mm'}))\\
&=f_{\mu_1}^{l_1 l_1'} \otimes f_{\mu_2}^{l_2 l_2'} \left(\sum_s \pi^\lambda_{ms} \otimes \pi^\lambda_{sm'}\right) = \sum_s f_{\mu_1}^{l_1 l_1'}(\pi^\lambda_{ms}) f_{\mu_2}^{l_2 l_2'}(\pi^\lambda_{sm'})\\
&= \sum_s \delta_{\mu_1 \lambda}\delta_{l_1 m} \delta_{l_1' s} \delta_{\mu_2 \lambda} \delta_{l_2 s} \delta_{l_2' m'}  = \delta_{\mu_1 \mu_2} \delta_{l_1' l_2} \delta_{\mu_2 \lambda} \delta_{l_1 m} \delta_{l_2' m'} \\
&=\delta_{\mu_1 \mu_2}\delta_{l_1' l_2} f_{\mu_2}^{l_1 l_2'}(\pi^\lambda_{m m'}) 
\end{split} 
\end{equation}
\begin{equation}
\begin{split}
&\;\Delta(f_\mu^{ll'})(\pi^{\lambda_1}_{m_1 m_1'}\otimes \pi^{\lambda_2}_{m_2 m_2'}) = f_\mu^{ll'}(\pi^{\lambda_1}_{m_1 m_1'}\times \pi^{\lambda_2}_{m_2 m_2'}) \\
&=\sum_\lambda C^{\lambda_1 \lambda_2 \lambda}_{m_1, m_2, m_1 + m_2} C^{\lambda_1 \lambda_2 \lambda}_{m_1', m_2', m_1' + m_2'}f_\mu^{ll'}(\pi^\lambda_{m_1+m_2, m_1'+m_2'})\\
&=\sum_\lambda C^{\lambda_1 \lambda_2 \lambda}_{m_1, m_2, m_1 + m_2} C^{\lambda_1 \lambda_2 \lambda}_{m_1', m_2', m_1' + m_2'}\delta_{\mu\lambda}\delta_{l, m_1+m_2}\delta_{l', m_1'+ m_2'} \\
&= C^{\lambda_1 \lambda_2 \mu}_{m_1, m_2, m_1 + m_2} C^{\lambda_1 \lambda_2 \mu}_{m_1', m_2', m_1' + m_2'}\delta_{l, m_1+m_2}\delta_{l', m_1'+ m_2'}  \\
&=\sum C^{\mu_1 \mu_2 \mu}_{l_1, l_2, l} C^{\mu_1 \mu_2 \mu}_{l_1', l_2', l'}\delta_{l, l_1+l_2}\delta_{l', l_1'+ l_2'}\delta_{\mu_1 \lambda_1}\delta_{l_1 m_1} \delta_{l_1' m_1'} \delta_{\mu_2 \lambda_2}\delta_{l_2 m_2}\delta_{l_2'm_2'}\\
&=\sum C^{\mu_1 \mu_2 \mu}_{l_1, l_2, l} C^{\mu_1 \mu_2 \mu}_{l_1', l_2', l'}\delta_{l, l_1+l_2}\delta_{l', l_1'+ l_2'}f_{\mu_1}^{l_1 l_1'}(\pi^{\lambda_1}_{m_1 m_1'})f_{\mu_2}^{l_2 l_2'}(\pi^{\lambda_2}_{m_1 m_1'}) \\
&=\sum C^{\mu_1 \mu_2 \mu}_{l_1, l_2, l} C^{\mu_1 \mu_2 \mu}_{l_1', l_2', l'}\delta_{l, l_1+l_2}\delta_{l', l_1'+ l_2'}f_{\mu_1}^{l_1 l_1'}\otimes f_{\mu_2}^{l_2 l_2'}(\pi^{\lambda_1}_{m_1 m_1'} \otimes \pi^{\lambda_2}_{m_2 m_2'})
\end{split}
\end{equation}
The product and coproduct on $F^*(H^*)$ are thus completely defined by:
\begin{gather}
\label{eqn:proddualF}
f_{\mu_1}^{l_1 l_1'}\times f_{\mu_2}^{l_2 l_2'} = \delta_{\mu_1 \mu_2}\delta_{l_1' l_2} f_{\mu_2}^{l_1 l_2'} \\
\Delta(f_\mu^{ll'})=  \sum C^{\mu_1 \mu_2 \mu}_{l_1, l_2, l} C^{\mu_1 \mu_2 \mu}_{l_1', l_2', l'}\delta_{l, l_1+l_2}\delta_{l', l_1'+ l_2'}f_{\mu_1}^{l_1 l_1'}\otimes f_{\mu_2}^{l_2 l_2'},
\end{gather}
where the sum in the last line is over $l_1,l_2,l_1',l_2',\mu_1$ and $\mu_2$. Note that this sum has an infinite number of non-vanishing terms corresponding to the pairs of irreducible representations $(\mu_1, \mu_2)$ whose tensor product contains the irreducible representation of $H^*$ labelled by $\mu$.\\
One can easily check that the unit and co-unit of $F^*(H^*)$ are given by:
\begin{gather}
1 = \sum_{\mu,l} f_\mu^{ll} \\
\epsilon(f_\mu^{ll'}) = \delta_{\mu0}\delta_{l0}\delta_{l'0}.
\end{gather}
Just as the coproduct, the unit is is not properly defined because it is a sums over an infinite number of basis elements with non-vanishing coefficients. This is just a formal problem because the finite dimensional representations of $F^*(H^*)$ will only pick out a finite number of elements as we shall see below.\\
\\
We can now truncate $F^*(H^*)$ to $F^*_+(H^*)$ by projecting out all functionals $f_{\mu}^{m m'}$ that do not satisfy the Murray condition for $G \to H$. This means that we will project out all functionals with $\mu$ not in the Murray cone $\Lambda_+$. The Murray condition can thus be implemented by using the following linear projection operator:
\begin{equation}
P: f_{ \mu}^{mm'} \mapsto P(f_{\mu}^{mm'}) = \left\{ \begin{array}{ll}
0 & \text{if $\mu \notin \Lambda_+ $} \\
f_{ \mu}^{mm'} & \text{if $\mu \in \Lambda_+$} \end{array}\right . 
\end{equation}
The product and coproduct of this truncated bi-algebra are given by: 
\begin{gather}
\label{eqn:prodproj}
f_{ \mu_1}^{l_1 l_1'}\times f_{ \mu_2}^{l_2 l_2'} = \delta_{\mu_1 \mu_2}\delta_{l_1' l_2} f_{ \mu_2}^{l_1 l_2'} \\
\Delta(f_{ \mu}^{ll'})=  \sum C^{\mu_1 \mu_2 \mu}_{l_1, l_2, l} C^{\mu_1 \mu_2 \mu}_{l_1', l_2', l'}\delta_{l, l_1+l_2}\delta_{l', l_1'+ l_2'}P(f_{ \mu_1}^{l_1 l_1'})\otimes P(f_{\mu_2}^{l_2 l_2'}). 
\end{gather}
Similarly, we have for the unit and co-unit:
\begin{gather}
1 = \sum_{\mu,l} P(f_{\mu}^{ll}) \\
\epsilon(f_{\mu}^{ll'})=\delta_{\mu0}\delta_{l0}\delta_{l'0}.
\end{gather}
\\
We shall now turn to the representations of $F^*(H^*)$ and $F^*_+(H)$. First we will introduce the  set $\{\pi^\lambda\}$ of representations of $F^*(H^*)$, where $\{\lambda\}$ is the set of irreducible representations of $H^*$.  We define these representations of $F^*(H^*)$ by:
\begin{equation}
\label{eqn:ddoublerep1}
\pi^\lambda: f_\mu^{ll'} \mapsto \pi^\lambda(f_\mu^{ll'}),
\end{equation}
where the matrix entries are given by
\begin{equation}
\label{eqn:ddoublerep2}
\left(\pi^\lambda(f_\mu^{ll'})\right)_{mm'} = f_\mu^{ll'}(\pi^\lambda_{mm'}).
\end{equation}
This definition ensures that the representations $\{\pi^\lambda\}$ respect the product of $F^*(H^*)$. 
The representations $\pi^\lambda$ defined here can thus be identified with the irreducible representations of  $H^*$. Below we shall prove that $\pi^\lambda$  itself is actually an irreducible representation of $F^*(H^*)$ and moreover we will find that these representations constitute the full set of irreducible representation of $F^*(H^*)$.\\
\\
Next we want to consider if the representations $\pi^\lambda$  of $F^*(H^*)$ are also representations of the truncated algebra $F^*_+(H^*)$. For $F^*(H^*)$ the representations above all satisfy 
\begin{equation}
\pi^{\lambda}(1)= \pi^{\lambda}\left(\sum f_{ \mu}^{ll}\right) = \mathbb I.
\end{equation}
However, in $F^*_+(H^*)$ we find for the representations labelled by $\lambda$ not satisfying the Murray condition: 
\begin{equation}
\pi^{ \lambda}( 1) = \pi^{\lambda}\left( \sum P(f_{ \mu}^{ ll})\right) =0.
\end{equation}
Since such $\pi^\lambda$ does not respect the identity this is not a representation of $F^*_+(H^*)$. Hence for $F^*_+(H^*)$ we must restrict to the truncated set of representations $\{\pi^{\lambda}\}$ satisfying the Murray condition. This last set of representations  is obviously in one-to-one relation with the magnetic charges in the fundamental Murray cone for $G \to H$. Below we shall prove that this is the complete set of irreducible representations of $F^*_+(H^*)$.\\
\\
We will construct the irreducible representations of $F^*(H^*)$ and $F^*_+(H^*)$ out of the representations of a set of subalgebras. Let $F^*$ denote either $F^*(H^*)$ or $F^*_+(H^*)$.  The subalgebras denoted by $F^*_\lambda\subset F^*$ are generated by $\{f_\lambda^{ll'}\}$ with fixed dominant integral weight $\lambda$. In the case of $F^*_+(H^*)$ we of course restrict $\lambda$ to be a dominant integral weight in $\Lambda_+$. Note that $\cup_\lambda F^*_\lambda = F^*$. It follows from the product rule (\ref{eqn:proddualF}) or (\ref{eqn:prodproj}) that $F^*_\lambda$ is indeed closed under multiplication. The identity $1_\lambda$ in $F^*_\lambda$ is expressed as:
\begin{equation}
1_\lambda=\sum_{l}f_{\lambda}^{ll}.
\end{equation}
These elements $1_\lambda \in F^*$ satisfy:
\begin{gather}
\label{eqn:communit}
f \times 1_\lambda = 1_\lambda \times f \quad \forall f \in F^*\\
\label{eqn:sumunit}
\sum_{\lambda}1_\lambda=1_{F^*} \\
\label{eqn:mutualanhil}
1_{\lambda}\times 1_{\lambda'} = \delta_{\lambda \lambda'} 1_{\lambda}. 
\end{gather}
We can use these properties to characterize the irreducible representations of $F^*$. Let $V$ be any irreducible representation of $F^*$. It is easy to see that for any $\lambda$ the image $V_\lambda$ of $V$ under the action of $1_\lambda$ is itself a representation of $F^*$. This follows from the fact that any $f\in F^*$ commutes with $1_\lambda$ as expressed by equation (\ref{eqn:communit}). $V$ thus contains invariant subspaces $\{V_\lambda\}$. For irreducible representations all invariant subspaces must be trivial, i.e.~equal either $\{0\}$ or $V$. Since any representation of $F^*$ respects the identity $1_{F^*}$ we find from (\ref{eqn:sumunit})  that for at least one $\lambda$ we must have $V_\lambda \neq \{0\}$, hence $V_\lambda=V$. Note that $\lambda$ is unique since  $V_{\lambda'}=\{0\}$ for $\lambda'\neq \lambda$ as follows from (\ref{eqn:mutualanhil}). Consequently any irreducible representation of $F^*$ is labelled by a dominant integral weight $\lambda$. It now follows from the product rule of $F^*$ that any $f_{\lambda'}^{ll'}\in F^*_{\lambda'}$ with $\lambda'\neq \lambda$ acts trivially on $V_\lambda$. An irreducible representation of $F^*$ thus corresponds to an irreducible representation of $F^*_\lambda$. Fortunately the irreducible representations of $F^*_\lambda$ are easily found.\\
\\  
Note that the labels $l$ and $l'$ of $F^*_\lambda$  take integer values in $\{1,\dots,n\}$ where $n$ is the dimension of the irreducible representation $\pi^\lambda$ of $H^*$. As it turns out $F^*_\lambda$ is a $n\times n$ matrix algebra and it is a well known fact that such an algebra has a unique irreducible representation of dimension $n$. For completeness we shall prove this now.\\ 
$F^*_\lambda$ has a commutative subalgebra ${F^*_\lambda}^{\text{diag}}$ generated by the elements $f_\lambda^{ll}$. Let us construct the irreducible representations of ${F^*_\lambda}^{\text{diag}}$. Since the algebra is commutative its irreducible representations are 1-dimensional. Let $\pi$ be such a representation. From $\pi(f_\lambda^{ll})^2 = \pi(f_\lambda^{ll}\times f_\lambda^{ll})= \pi(f_\lambda^{ll})$ we find that $\pi(f_\lambda^{ll})$ equals either 0 or 1. If we assume the latter for a fixed value $k$ of $l$ then we have for $l\neq k$: 
\begin{equation}
\pi(f_\lambda^{ll})= \pi(f_\lambda^{kk})\pi(f_\lambda^{ll}) = \pi(f_\lambda^{kk}\times f_\lambda^{ll})= \delta_{kl}\pi(f_\lambda^{ll})=0.
\end{equation}
Note that since the unit of ${F^*_\lambda}^{\text{diag}}$ must be respected $\pi(f_\lambda^{ll})$ cannot vanish for all $l$. The irreducible representations of ${F^*_\lambda}^{\text{diag}}$ are thus given by:
\begin{equation}
\pi_l: f_\lambda^{l'l'} \mapsto \delta_{ll'}.
\end{equation}
Any non-trivial irreducible representation $(\pi,V)$ of $F^*_\lambda$ can be decomposed into a sum of irreducible representations of ${F^*_\lambda}^{\text{diag}}$. Hence there is a  $v^k\in V$ such that  $\pi(f_{\lambda}^{ll})v^{k} = \delta_{l k} v^{k}$. Let us define a set of $n$ vectors in $V$ by $v^m=\pi(f_{\lambda}^{mk})v^k$. The span of $\{v^m\}$ defines an invariant subspace of $V$. This follows again from the product rule:
\begin{equation}
\label{eqn:diaginv}
\begin{split}
\pi(f_{\lambda}^{l l'})v^{m} &= \pi(f_\lambda^{l l'})\pi(f_\lambda^{m k}) v^k = \pi(f_\lambda^{l l'}\times f_\lambda^{m k}) v^k \\
                            &= \delta_{l' m}\pi(f_\lambda^{l k}) v^k = \delta_{l' m} v^{l}.
\end{split}
\end{equation}
Since $\pi$ is irreducible the span of $\{v^m\}$ is $V$.\\ 
\\
The claim is that $V$ is $n$-dimensional. In order to prove this we have to show that the vectors $v^m$ are linearly independent. If 
\begin{equation}
\sum_{m} a_m v^{m}=0 
\end{equation}
one finds from (\ref{eqn:diaginv}):
\begin{equation}
f_{\lambda}^{ll} \left(\sum_{m} a_m v^{m}\right) = a_l v^{l}=0.
\end{equation}
So either $a_l=0$ or $v^l=0$. However, $v^l=0$ together with the product rule and the definition of $v^m$ implies that $v^{m}=\pi(f_{\lambda}^{ml})v^{l}=0$. This would mean that $V=\{0\}$ contradicting the fact that $V$ is non-trivial i.e.~at least one dimensional. We thus find that an irreducible representation of $F^*_\lambda$ is $n$-dimensional and moreover it follows from the explicit action on a basis of $V$ as in equation (\ref{eqn:diaginv}) that such an irreducible representation is unique up to isomorphy.\\
\\
We have found that an irreducible representation of $F^*$ is completely fixed by a dominant integral weight $\lambda$ in the appropriate weight lattice. The dimension of such an irreducible representation is given by the dimension of the irreducible representation of $H^*$ with highest weight $\lambda$. To find the fusion rules for these representations we go back to the representations $\{\pi^\lambda\}$ introduced in formula (\ref{eqn:ddoublerep1}) and (\ref{eqn:ddoublerep2}) via the matrix entries of the original $H^*$-representations. The dimensions of these representations are given by the dimensions of the corresponding highest weight representations of $H^*$. Moreover they satisfy $\pi^\lambda(f_\mu^{ll'})=0$ for $\mu\neq \lambda$, i.e.~$\pi^\lambda$ defines a representation of $F_\lambda^*$.  By comparing (\ref{eqn:ddoublerep2}) and (\ref{eqn:diaginv}) one finds that $\pi^\lambda$ corresponds precisely to the unique non-trivial irreducible representation of $F^*_\lambda$. We conclude that the representations $\{\pi^\lambda\}$ are the irreducible representations of $F^*$. Since the labels, the matrix elements and hence also the dimensions of these irreducible representations match those of the irreducible representations of $H^*$  is seems very likely that the fusion rules for these representations of $F^*$ are also identical to the fusion rules of the corresponding $H^*$-representations.\\
\\
We have seen that the representation of $F^*(H^*)$ are identical to the representations of $H^*$. One might thus wonder to what extent $H^*$ and $F^*(H)$ are equivalent. If $H^*$ is a finite group one would find that $F^*(H^*)$ being a double dual of $\mathbb C H^*$ is isomorphic to the group algebra $\mathbb CH^*$. Since in our cases $H^*$ is a continuous group one has to take care in taking the dual. Nonetheless one can define $F(H^*)$ as the dual of $H^*$ via the irreducible representations of $H^*$. Similarly, one can retrieve $H^*$ from the co-representations of $F^*(H)$. These co-representations are nothing but the representations of  $F(H)$  which is the dual of $F^*(H)$. Let us illustrate this for $H^*=U(1)$.\\
An irreducible representation of $U(1)$ is uniquely labelled by an integer number. It is not very hard to check from equation (\ref{eqn:dualprod}) that the product of $F(U(1))$ can be expressed as:   
\begin{equation}
\label{eqn:prodU1}
\pi^{n} \times \pi^{n'} =  \pi^{n+n'}.
\end{equation}
Since $F(U(1))$ is commutative its irreducible representations are 1-dimen\-sional. An irreducible representation thus sends $\pi^1$ to some $z\in \mathbb C$. It follows from (\ref{eqn:prodU1}) that the representation is completely defined by $z$:
\begin{equation}
z: \pi^n \mapsto z^n \in \mathbb C.
\end{equation}
Not all values of $z$ give a representation of $F(U(1))$ though. To give an example we note that $\pi^{-1}$ is mapped to $z^{-1}$. This goes wrong for $z=0$. For each $z\in \mathbb C\setminus\{0\}$ one does find a proper representation. It is easy to check that $\mathbb C\setminus\{0\}$ is a group.  Obviously this is not the group $U(1)$. As matter of fact we have reconstructed the complexification $U(1)_{\mathbb C}$ of $U(1)$. To understand this we note that $U(1)$ has an involution which takes $h\mapsto h^* = h^{-1}$. The representations of $U(1)$ respect this involution in the sense that $(\pi^n(h))^*=\pi^n(h^*)$. We therefore have a natural involution on $F^*(U(1))$ defined by $(\pi^n)^* = \pi^{-n}$. Again one can  define the representations of $F(U(1))$ to respect the involution, i.e.~$(z(\pi^n))^*= z((\pi^n)^*)$. This results in the condition $z^*=z^{-1}$ which restricts $z$ to the unit circle in $\mathbb C$, i.e.~to $U(1)$.\\
\\
For $SU(2)$ broken to $U(1)$ the Murray cone is the set of all non-negative integers. This implies that the dual $F_+(U(1))$ of  $F_+^*(U(1))$ is generated by $\{\pi^n : n\geq 0\}$. The product is still given by (\ref{eqn:prodU1}) and hence $F_+(U(1))$ is a commutative algebra. Again we define an irreducible representation by $\pi^1 \mapsto z\in \mathbb C$. Since the representation should respect the product we find that the choice of $z$ completely fixes the representation, i.e.~$z: \pi^n \mapsto z^n$. One might again wonder if all values of $z$ give a representation. Note that $F_+(U(1))$ is not closed under inversion just as the Murray cone is not closed under inversion. For example $\pi^{-1}\notin F_+(U(1))$. The representation labelled by $z=0$ is thus not immediately ruled out. Note that for $z=0$ we have $\pi^n\mapsto 0$ for all $n>0$. The image of $\pi^0$ seems undetermined, nonetheless we can set $z(\pi^0)=z_0 \in \mathbb C$ for $z=0$. The representation we now obtain does respect the product if and only if $z_0$ equals either $0$ or $1$. But since $\pi^0$ is the unit of the algebra it should be mapped to the unit of $\mathbb C$. Hence we find that $z(\pi^0)=1$ for all $z\in \mathbb C$ and in particular for $z=0$.\\
We have found that we should identify $U(1)_+$ with $\mathbb C$. The complex numbers are indeed closed under multiplication and moreover this multiplication is associative. On the other hand there is no inverse. We thus see that $U(1)_+$ is a semi-group and not a group as $U(1)$. Let us finally connect both ends of the circle and see if the commutative algebra  $U(1)_+=\mathbb C$ has the appropriate irreducible representations. Obviously $U(1)_+$ has representations $\pi^n$ for $n>0$ defined by:
\begin{equation}
\pi^n: z \mapsto z^n.
\end{equation} 
Representations with $n<0$ do not exist because the image of $z=0$ would not be defined. Finally the representation $\pi^0$ is a bit tricky. One can however simply define $\pi^0(0)=z_0$. It follows from the product on $\mathbb C$ that $z_0$ equals either $0$ or $1$. If however we restrict all representations to be continuous we find $z_0=1$.  It is now almost trivial to check that the fusion rules of $U(1)_+$ correspond precisely to the fusion rules of $U(1)$. It would be interesting to study if a smooth semi-group $H^*_+$ can be defined for every possible residual dual gauge group $H^*$.   
\newpage
\addcontentsline{toc}{section}{References}
\bibliography{monopolebib2}
\bibliographystyle{JHEP}
\end{document}